\documentclass[10pt]{article}
\usepackage{amsmath,amssymb,graphicx}
\batchmode

\newtheorem{lem}{Lemma}
\newtheorem{thm}{Theorem}

\newtheorem{cor}{Corollary}

\newcommand{\pr}{\noindent{\bf Proof}. }
\newcommand{\re}{\noindent{\bf Remark}. }
\newcommand{\res}{\noindent{\bf Remarks}. }

\newcommand{\pa}{\partial}

\newcommand{\bpsi}{\bar \psi}
\newcommand{\bPsi}{\bar \Psi}
\newcommand{\const}{\textrm{const }}
\newcommand{\st}{\stackrel{}}

\newcommand{\bom}{\mbox{\boldmath $\Om$}}
\newcommand{\bla}{\mbox{\boldmath $\La$}}
\newcommand{\bfpsi}{\mbox{\boldmath $\Psi$}}

\newcommand{\bfA}{\mbox{\boldmath $A$}}

\newcommand{\bfD}{\mbox{\boldmath $D$}}

\newcommand {\bfa}{\mbox{\boldmath $a$}}
\newcommand {\bfb}{\mbox{\boldmath $b$}}

\newcommand{\al}{\alpha}
\newcommand{\De}{\Delta}
\newcommand{\de}{\delta}
\newcommand{\ga}{\gamma}
\newcommand{\Ga}{\Gamma}

\newcommand{\La}{\Lambda}
\newcommand{\la}{\lambda}
\newcommand{\Om}{\Omega}
\newcommand{\om}{\omega}

\newcommand{\si}{\sigma}

\newcommand{\cA}{{\cal A}}

\newcommand{\cV}{{\cal V}}
\newcommand{\cO}{{\cal O}}

\newcommand{\cH}{{\cal H}}
\newcommand{\cS}{{\cal S}}
\newcommand{\cE}{{\cal E}}

\newcommand{\cK}{{\cal K}}

\newcommand{\cM}{{\cal M}}
\newcommand{\cN}{{\cal N}}

\newcommand{\cQ}{{\cal Q}}

\newcommand{\cJ}{{\cal J}}
\newcommand{\cZ}{{\cal Z}}

\newcommand{\bbR}{{\mathbb{R}}}
\newcommand{\bbZ}{{\mathbb{Z}}}

\newcommand{\bbT}{{\mathbb{T}}}
\newcommand{\bbD}{{\mathbb{D}}}

\begin{document}

\title{Quantum electrodynamics on the 3-torus \\ \small{II.- The RG flow}}
\author{ 
J. Dimock\thanks{Research supported by NSF Grant PHY0070905}
\thanks{dimock@acsu.buffalo.edu}\\
Dept. of Mathematics \\
SUNY at Buffalo \\
Buffalo, NY 14260 }
\maketitle

\begin{abstract}
We   continue the   study of  quantum electrodynamics  on a
three dimensional torus as the limit of a lattice gauge theory.    
In this paper we give a preliminary treatment of the renormalization
group   flow.  We study the propagators which arise under
multiple block spin averaging, both in global and local versions.   We also study low order
perturbation theory.  However we  do not control remainders.  This 
is left to the more complete treatment of the following paper.
\end{abstract}

\newpage

\tableofcontents

\newpage
\section{Global flow}  \label{one}

\subsection{overview}

In this paper we continue our analysis of QED on the  3-torus which was begun in  paper I \cite{Dim02}.
In that paper we   explained the first step  consisting of a a renormalization group (RG)
transformation  followed by   a split into regions with  large and small gauge fields.  
In this paper we study multiple RG transformations as much as possible \textit{without} the split
into regions of large and small fields.  The emphasis is on studying the propagators which 
arise under multiple block spin averaging,  both in global and localized versions.   We also 
study low order perturbation theory, but as yet without estimating remainders.
The paper is preliminary to   paper  III  \cite{Dim05}  in which we also incorporate the
split into large and small fields at each step.    This gives  the full
expansion and gives an   expression for the effective action on any length scale. The expression exhibits
the contribution of low order perturbation theory and controls the remainder.  
 
The analysis starts with a field theory on a toroidal lattice  $\bbT^{-N}_M = L^{-N}\bbZ^3/ L^M\bbZ^3 $ with
spacing  $L^{-N}$  and volume  $L^{3M}$.  Our goal is to gain some control over the 
$N \to \infty$ limit for fixed $M$.   We start by scaling the theory up to  $\bbT^{0}_{N+M}$
which has unit spacing and volume  $L^{3(M+N)}$.  The density for the theory 
is  then    
\begin{equation}  \label{rho0}
\begin{split}
 \rho_0(\Psi_0 , A_0)  
=\exp \left( - \frac12  (A_0, (- \De + \mu_0^2)A_0) 
- ( \bPsi_0, ( D_{e_0}(A)+ m_0 ) \Psi_0)  - v_0 -  \cE_0 \right)
\end{split}
\end{equation}
Here the gauge fields $A_0= \{A_{0,\mu}(x) \}$ are real numbers indexed by  $x \in\bbT^{0}_{N+M}$ 
and   $\mu=(1,2,3)$.  The fermi fields   
  $\Psi_0  =   \{ \Psi_{0,\al}(x)\}$ 
and     $\bPsi_0  =   \{  \bPsi_{0,\al}(x)\}$ 
are the generators of a Grassmann algebra and are indexed   by  $x \in  \bbT^{0}_{N+M}$
and spinor indices  $ \al =(1,\dots,4)$.
The operator $\De $ is the lattice Laplacian and  $ D_{e_0}(A)$
is the lattice Dirac operator with gauge field $A$.
The coupling constant $e_0$, fermion mass $m_0$, and boson mass $\mu_0$ 
are  scaled versions of their bare values  $e,m,\mu$ :
\begin{equation}   e_0 = L^{-N/2} e  \ \ \ \ \ m_0  = L^{-N}m
\ \ \ \ \ \mu_0 = L^{-N}\mu
\end{equation}
Finally  $ v_0$ is a fermion mass counterterm and  $\cE_0$ is a vacuum energy counterterm.
The counterterms play no role in the present paper and are set equal to zero.   
See  paper I for more details.

To  analyze this expression we perform a number of renormalization group  (RG) transformations 
  consisting  of averaging operation    followed by scalings back to the unit lattice. 
We get a sequence of  densities  $\rho_0,\rho_1, \rho_2, \dots  $.
The  density    $\rho_k$ is   defined on fields   $\Psi_k, A_k$  on   $\bbT^{0}_{N+M-k}$.
The sequence is generated by  
\begin{equation}   \label{firstrep}
\begin{split}
 & \rho_{k+1}(\Psi_{k+1}, A_{k+1}) \\ = &
\int  \cN_{k+1,a}^{-1}  \cM_{k+1,b}^{-1}
   \exp \left(
- \frac12 \frac{a}{L^2}  |A_{k+1,L}- Q A_{k}|^2   
 - \frac{b}{L}   |\Psi_{k+1,L}- Q_{e_k}( A'_k) \Psi_k|^2  
\right)  \rho_k(\Psi_k , \cA_k) \    dA_k \ d\Psi_k
\\
\end{split}
\end{equation} 
Here scaling up  by  $L$ is defined by  
\begin{equation}
\begin{split}
  A_L(x)  = &   (\si^b_LA)(x) = L^{-1/2}A(L^{-1}x) \\
\Psi_L(x)   = & (\si^f_L  \Psi)(x) = L^{-1} \Psi(L^{-1} x ) \\    
\end{split}
\end{equation}
so that   $\Psi_{k+1,L}, A_{k+1,L}$ are   fields  on  $\bbT^{1}_{N+M-k}$ 
The operators $Q,Q_{e_k}(A_k')$  average over blocks of side L and also
  take us to fields on  $\bbT^{1}_{N+M-k}$.  
For  fermions  this has the gauge covariant form    
\begin{equation}
(Q_{e_k}(A_k')\Psi)(y)    =  L^{-3}  \sum_{|x-y| \leq L/2}\exp \left( ie_k 
 A'_k(\Ga_{yx}) \right) \Psi(x)
\end{equation}
Here    $e_k = L^{-(N-k)/2} e $ 
is   a running coupling constant,   the gauge field  $A_k'$  is either  $A_k$
or some modification to be specified,  and  $\Ga_{yx}$ is a standard path from $x$ to $y$.  We are using the
notation
\begin{equation}
 |\Psi_{k+1,L}- Q_{e_k}( A'_k) \Psi_k|^2 =  (\bPsi_{k+1,L}- Q_{e_k}(- A'_k) \bPsi_k,   \Psi_{k+1,L}-
Q_{e_k}( A'_k) \Psi_k)
\end{equation}
The normalization constants are chosen  so that  
\begin{equation}
  \int  \rho_{k+1}(\Psi_{k+1}, A_{k+1})   d\Psi_{k+1} dA_{k+1} =  \int  \rho_{k}(\Psi_k, A_k)   d\Psi_{k}
dA_{k}
\end{equation}
That is 
\begin{equation}
\begin{split}
 \cN_{k,a}^{-1}  = & \int  e^{-a |A_{k}|^2 /2}  dA_{k} 
=  \left( \frac {2 \pi}{a} \right)^{3|\bbT^0_{N+M-k}|/2}  \\
 \cM_{k,b}^{-1}
= &
 \int    e^{  - b ( \bPsi_{k},  \Psi_{k})  }  d\Psi_{k}
= b^{4|\bbT^0_{N+M-k}|}  \\
\end{split}
\end{equation}
We want to study $\rho_k$ as  $ k \to N$.

Before proceeding we recall our indebtedness to  the papers of Balaban and collaborators: 
\cite{Bal82a}  - \cite{Imb86}.

\subsection{bosons}
\subsubsection{}
Let us start  by dropping the fermions and
study the effect of repeated RG transformations for the bosons.
If we do the intermediate integrals and scale down we find the 
the following formula
 \begin{equation}   \label{combine}
\begin{split}
&\int\prod_{j=0}^{k-1}dA_j \cN_{j+1,a}^{-1}   \exp \left(
- \frac12 \frac{a}{L^2}  |A_{j+1,L}- Q A_{j}|^2  \right) 
 \exp \left(- \frac12  ( A_0,(-\De+ \mu^2_0) A_0) \right) F(A_0) \\
=&\int   \cN_{k,a_k}^{-1}      \exp \left(
- \frac{ a_k}{2}  |A_{k}- \cQ_k \cA|^2   - \frac12  ( \cA,(-\De+ \mu^2_k) \cA)
\right)  F(\cA_{L^k}) d\cA\\
\end{split}
\end{equation}
Here  we are integrating over functions  $\cA$ on    $\bbT^{-k}_{N+M-k}$
and the inner product is 
\begin{equation}
(\cA, \cA)  =  \sum_{x }  L^{-3k} |\cA(x)|^2  \equiv
\int |\cA(x)|^2 dx
\end{equation}
The operator   $\cQ_k$ is an averaging operator on this space taking us
to functions on  the unit lattice  $\bbT^{0}_{N+M-k}$  given by  
\begin{equation}
(\cQ_k \cA)_{\mu}(y)  = \int_{|x-y| \leq 1/2} \cA_{\mu}(x) dx
\end{equation}
With  $\sigma_L  = \sigma^b_L$  it   satisfies 
\begin{equation}  \label{bQ}
\cQ_{k+1}
=  \si_{L^{-1}}   Q \ \cQ_{k}  \ \si_{L}
\end{equation}
The numbers  $a_k$ obey the recursion equation   $a_{k+1}  = a a_k/ ( a_k + a/L^2)$
which has the solution
\begin{equation}
a_k  = \left( \frac{1-L^{-2}}{1-L^{-2k}}  \right) a
\end{equation}
The quadratic form $( \cA,(-\De+ \mu^2_k) \cA)$ is the scaled version of  
$( A_0,(-\De+ \mu^2_0) A_0)$ and has mass  $\mu_k  = L^{-(N-k)}\mu $.  

  The formula (\ref{combine}) is proved by induction.
 It comes down to the  explicit calculation of the Gaussian integral  
\begin{equation}  \label{combine2}
\begin{split}
&\int  d A_k \  \cN_{k+1,a}^{-1}     \cN_{k,a_k}^{-1}     
 \exp \left( - \frac{a}{2L^2}   |A_{k+1,L}- Q A_k|^2  \right) 
 \exp \left( -  \frac{a_k}{2}   |A_{k}- \cQ_{k} \cA_L|^2  \right)  \\
\ \ \ \ \ \ \ \ = &   \cN_{k+1,a_{k+1}}^{-1}    \exp \left(
 - \frac{a_{k+1}}{2}   |A_{k+1}- \cQ_{k+1} \cA|^2  
\right) \\
\end{split}
\end{equation}
which we carry out in appendix  \ref{A}.

\subsubsection{}
To evaluate integrals such as those on the right side of  (\ref{combine})
we use the identity 
\begin{equation}  \label{Gauss}
\begin{split}
&  \int   \cN_{k,a_k}^{-1}   \exp \left(
- \frac{ a_k}{2}  |A_{k}- \cQ_k \cA|^2   - \frac12  ( \cA,(-\De+ \mu^2_k) \cA)
\right)  f(\cA) d\cA\\
=     & Z_k \ \exp \left(- \frac 12 (A_k,  \De_k A_k)\right) 
\int   f( \cA+\cH_k A_k )\  d \mu_{G_k} (\cA)\\
\end{split}
\end{equation}
Here we have defined    
\begin{equation}
\De_k^\# = -\De+ \mu^2_k  +  a_k  \cQ_k^T \cQ_k 
\end{equation}
and   
\begin{equation}
\begin{split}
G_k =&  (\De_k^{\#} )^{-1}\\
\cH_k =&  a_k  (\De_k^{\#} )^{-1} \cQ_k^T \\
\De_k   =&  a_k \ I  -  a_k^2  \cQ_k  (\De_k^{\#} )^{-1} \cQ_k^T  \\
\end{split}
\end{equation}
The operators $\De_k^\#, G_k$ act on functions on  $\bbT^{-k}_{N+M-k}$, 
$\cH_k $ maps functions on  $\bbT^{0}_{N+M-k}$  to functions on   $\bbT^{-k}_{N+M-k}$, 
and   $\De_k$  is an operator on functions on   $\bbT^{0}_{N+M-k}$ . 
We have also defined    $ \mu_{G_k}$ to  be the Gaussian measure with covariance  $G_k$.

To prove (\ref{Gauss}) 
make transformation  $\cA\to  \cA  +  \cH_k A_k$. This  diagonalizes the quadratic form
and gives   $- \frac 12 (A_k,  \De_k A_k) - \frac12  ( \cA,\De_k^\# \cA)$, whence 
the result with  
\begin{equation}
Z_k  =   \cN_{k,a_k}^{-1}   
\int e^{ - \frac12  ( \cA,(-\De_k^\#) \cA)}d\cA
\end{equation}

\subsubsection{}
 As a special case  we  take   $f(\cA)  =   \exp(i(\cA,J))$ and       consider  the generating function
\begin{equation}\Om_k(A_k,J)=
\int  \cN_{k,a_k}^{-1} \exp \left(
- \frac{ a_k}{2}  |A_{k}- \cQ_k \cA|^2   - \frac12  ( \cA,(-\De+ \mu^2_k) \cA)+i (J,\cA)
\right)  d\cA
\end{equation}
Using    (\ref{Gauss}) and $\int \exp(i(\cA,J)) d \mu_{G_k}(\cA) =  \exp( - 1/2 (J,G_kJ))$       this is
evaluated as   
\begin{equation}  \label{generate}
 \Om_k(A_k,J)=
Z_k  \exp \left(- \frac 12 (A_k,  \De_k A_k)) - \frac12  (J,G_kJ)  +i(J,\cH_k A_k) \right)
\end{equation}

We can also take it one step at a time.  Using (\ref{combine2}) we have    
\begin{equation}   \label{onestep}
\Om_{k+1}(A_{k+1},J)  = \int  \cN_{k+1,a}^{-1}  \exp \left(
- \frac12 \frac{a}{L^2}  |A_{k+1,L}- Q A_{k}|^2   \right)   \Om_k(A_k,   \si_{L^{-1}}^T J)
dA_k
\end{equation}
To evaluate we  introduce 
\begin{equation}
\begin{split}
C_k =&  ( \De_k   + \frac{ a}{L^2}  Q^TQ )^{-1}\\
H_{k} =& \frac{a}{L^2}  C_k Q^T \\
\end{split}
\end{equation}
Now insert  (\ref{generate}) into    (\ref{onestep}) and make  the transformation  $A_k \to   A_k +   H_{k}
A_{k+1,L}$ to diagonalize the quadratic form.
The integral over $A_k$ becomes a Gaussian integral with covariance  $C_k$.  Carry out the 
integral and compare the resulting expression with  (\ref{generate}) for  $\Om_{k+1}(A_{k+1},J) $.
 We find that  
  \begin{equation}  \label{iterate}
\begin{split}
G_{k+1}   = & \si_L^{-1} (  G_k  +   \cH_k  C_k \cH_k^T  )  (\si_L^{-1})^T \\
\cH_{k+1}   = &  \si_L^{-1}  ( \cH_k  H_{k}) \si_L  \\
\De_{k+1}   =& \si_L^T   \left(\frac{a}{L^2}  -  \frac{a^2}{L^4}  Q C_k Q^T\right)\si_L   \\
\end{split}
\end{equation}
(for  $k=0$ the conventions are  $G_0=0$,  $\cH_0 =I$, and $\De_0 =  -\De + \mu_0^2$ )
We also have  
\begin{equation}  \label{zz}
 Z_{k+1}  =
Z_k \ \cN_{k+1,a}^{-1} \int  e^{- \frac12 (A_k, C_k^{-1} A_k)} d A_k
\end{equation}

\subsubsection{}

As an operator on   functions on $\bbT^{-k}_{N+M-k}$, the propagator $G_k$ 
has a kernel   $G_{k}(x,x')$  defined so that 
$(G_k f)(x)  =\int   G_{k}(x,x') f(x') dx'  $  
where again the integral means the weighted sum.
Since   $\si_{L^{-1}}^T =  L^{-2} \si_L$ for bosons and    $(\si_L^{-1} G_k  \si_L)(x,x')  =  L^3
G_k(Lx,Lx')$ and  we can write (\ref{iterate}) as   
\begin{equation}
G_{k+1} (x,x')  = 
 L ( G_k(Lx,Lx')  +  ( \cH_k  C_k \cH_k^T)(Lx,Lx') )
\end{equation}
If we iterate this we find
\begin{equation}  \label{decompb}
G_k(x,x')  
= 
\sum_{j=0}^{k-1}L^{k-j} \tilde C_j(L^{k-j}x, L^{k-j}x') 
\end{equation}
where  
\begin{equation}
\tilde C_j  = \cH_jC_j \cH_j^T 
\end{equation}

The operators  $ \cH_j, C_j, \tilde  C_j$  have    kernels  which satisfy 
for  $x,x' \in \bbT^{-k}_{N+M-k}$ and   $y,y' \in \bbT^{0}_{N+M-k}$
\footnote{In our notation $\cO(1)$ allows $L$ dependence.  We do however note that 
in the second  and third bounds  the  $\cO(1)$ in the
exponent is  actually  $\cO(L^{-1})$.}
\begin{equation}
\begin{split}
|\cH_j( x,y)|,  |\pa \cH_j( x,y)|,    \leq & \cO(1) \exp( - \cO(1) d(x,y) )\\
|C_j(y,y')|  \leq &  \cO(1) \exp( - \cO(1) d(y,y') )\\
|\tilde C_j(x,x')|  \leq &  \cO(1) \exp( - \cO(1) d(x,x')) \\
 \end{split}
\end{equation}
These are proved by Balaban  in \cite{Bal83b}, but one may prefer to use the methods of  \cite{BOS89}.

Using this in  (\ref{decompb}) leads to the estimates 
\begin{equation}  \label{uno}
\begin{split}
| G_k(x,x')| \leq  & 
 \left\{ \begin{array}{rl}
\cO(1) d(x,x')^{-1} e^{- \cO(1) d(x, x'  )}  &  x\neq x' \\
\cO(L^k )  & x=x'\\  
\end{array}  \right.     
\\  
|\pa G_k(x,x')| \leq  & 
 \left\{ \begin{array}{rl}
\cO(1) d(x,x')^{-2} e^{- \cO(1) d(x, x'  )}  &  x\neq x' \\
\cO(L^{2k} )  & x=x'\\  
\end{array}  \right.     
\\  
\end{split}
\end{equation}
(To see the short distance bound divide the sum into terms satisfying  $L^{k-j} d(x,x') \leq  1$
and the complement.)

The function  $G_k(x,x')$ is our basic photon propagator after $k$
steps.  This estimate shows the exponential decay whose origin
is an effective mass from  $\cQ_k^T\cQ_k$, and the characteristic
short distant singularity  $d(x,x')^{-1} $.

\subsection{fermions}

\subsubsection{}
Next we want to do something similar for fermions.
For an  arbitrary element   $ F(\Psi_0)$ in the Grassmann algebra generated by 
$\Psi_0$   we  claim that   
\begin{equation}  \label{old}
\begin{split} 
& \int    \prod_{j=0}^{k-1} d \Psi_j \cM_{j+1,b} ^{-1} \exp \left(
 - \frac{b}{L}   |\Psi_{j+1,L}- Q_{e_j}( \tilde \cQ_j \cA_j) \Psi_j|^2  
\right)   \exp \left( - (\bPsi_0,(D_{e_0}(\cA_0)+m_0) \Psi_0)
\right) \ F(\Psi_0) \\
=& \int d \psi  \cM_{k,b_k} ^{-1}  \exp \left(
 - b_k   |\Psi_{k}- \cQ_{k}(\cA_{k-1,L^{-1}}, \dots , \cA_{0,L^{-k}}) \psi|^2  
\right) \\
&\ \ \ \ \ \ \ \ \ \ \ \ \ \ \ \exp \left( - (\bpsi,(D_{e_k}(\cA_{0,L^{-k}})+m_k) \psi) 
\right)   F(\psi_{L^k}) \\
\end{split}
\end{equation}
Here on the left side instead of  $ Q_{e_j}(  A'_j) $ we have taken 
$ Q_{e_j}( \tilde \cQ_j \cA_j)$ 
where    $\cA_j$ is for the moment an arbitrary function on  $\bbT^{-j}_{N+M-j}$
 and  $ \tilde \cQ_j $ is an averaging operator which brings the field  up  to  
$\bbT^0_{N+M-j}$.   The averaging operator   is not the same as  $\cQ_j$, but given by 
\begin{equation}
(\tilde \cQ_j\cA)_{\mu}(y)  = 
\int_{|x-y| \leq 1/2} \cA(\Ga_{yx} \cup [x, x+ e_{\mu}] \cup \Ga_{x+e_{\mu},y+ e_{\mu}})  
\end{equation}
where $\cA(\Ga)  $ indicates the integral along the contour 
 $\Ga$.   For  a scalar  $\la$ on  $\bbT^{-j}_{N+M-j}$      
\begin{equation}
(\tilde \cQ_j (\cA + d\la))_{\mu}(y) =(\tilde \cQ_j\cA)_{\mu}(y)  +  \la(y) -\la(y+e_{\mu})
\end{equation}
 so this averaging is gauge covariant.  

On the right side of  (\ref{old}) we are integrating over  $\psi$ on   $\bbT^{-k}_{N+M-k}$.
 The multiple    averaging operators $ \cQ_{k}(a_{k-1}, \dots , a_{0})$ 
depend on fields   $a_{k-1}, \dots , a_{0}$ all on   $\bbT^{-k}_{N+M-k}$.
They are defined recursively 
by  
\begin{equation}    \label{together}
\cQ_{k+1}(a_{k}, \dots , a_0) 
=  \si_{L^{-1}}\   Q_{e_k}(\tilde \cQ_k a_{k,L})\
 \cQ_{k}(a_{k-1,L}, \dots , a_{0,L})  \ \si_{L}
\end{equation}
where  $a_{k}, \dots , a_{0}$ are  all on   $\bbT^{-k-1}_{N+M-k-1}$ and  $\si_L  =  \si^f_L$.
An explicit expression is given in Appendix \ref{A}.
The numbers $b_k$ obey 
\begin{equation}
b_k  = \left(
\frac{1-L^{-1}}{1-L^{-k}} 
\right) b
\end{equation}
The form  $(\bpsi,(D_{e_k}(\cA)+m_k) \psi)$ has the Dirac operator on    $\bbT^{-k}_{N+M-k}$
and mass $m_k =  L^{-(N-k)}m$.

The formula  (\ref{old}) is again   proved by induction.   It comes down to the
Gaussian integral   
\begin{equation}  \label{newt}
\begin{split}
 &\int  d \Psi_k  \cM_{k+1,b}^{-1}     \cM_{k,b_k}^{-1} 
 \exp \left(
 - \frac{b}{L}   |\Psi_{k+1,L}- Q_{e_k}( \tilde \cQ_k \cA_k) \Psi_k|^2  
\right)  \\
&\ \ \ \ \ \ \ \ \ \ 
 \exp \left(
 - b_k   |\Psi_{k}- \cQ_{k}(\cA_{k-1,L^{-1}}, \dots , \cA_{0,L^{-k}}) \psi_L|^2  
\right)  \\
&=\cM_{k+1,b_{k+1}}^{-1}  \exp \left(
 - b_{k+1}   |\Psi_{k+1}- \cQ_{k+1}(\cA_{k,L^{-1}}, \dots , \cA_{0,L^{-k-1}} )\psi|^2  
\right) \\ 
\end{split}
\end{equation}
which we prove in Appendix \ref{A}.

\subsubsection{}
We specialize to the  case where all the fields are scalings of the last.  For
any $\cA$ on  $\bbT^{-k}_{N+M-k}$, let   $\cA_j = \cA_{L^{k-j}}$.  Then on the right
side of (\ref{old}) we identify  
\begin{equation}
\cQ_k(\cA) \equiv    \cQ_{k}(\cA, \dots , \cA)
\end{equation}
From appendix \ref{A}
\begin{equation}  \label{explicit}
\begin{split}
(\cQ_k(\cA) \psi)(y)  =&  \int_{|x-y| <1/2}  \exp( ie_k \cA(\tilde \Ga_{yx}) ) \psi(x) dx  \\
\cA(\tilde \Ga_{yx}) = & \sum_{j=0}^{k-1} (\tilde \cQ_j \cA)(\Ga_{x_{j+1}, x_j})  \\
\end{split}
\end{equation}

Now assume that  $e_k | \pa \cA| $ 
is sufficiently small.
 Then we claim that an integral like   the right side of    
(\ref{old})  with   $\cQ_k(\cA)$ can be evaluated as  
\begin{equation}   \label{new}
\begin{split}
& \int d \psi  \cM_{k,b_k} ^{-1}  \exp \left(
 - b_k   |\Psi_{k}- \cQ_{k}( \cA) \psi|^2  
\right) \exp \left( - (\bpsi,(D_{e_k}(\cA)+m_k) \psi) 
\right)   f(\psi) \\
&=      Z_k(\cA) \ \exp \left(-  (\bPsi_k,  D_k(\cA) \Psi_k)\right) 
\int   f( \psi+\cH_k(\cA) \Psi _k )\  d \mu_{S_k(\cA)} (\psi)\\
\end{split}
\end{equation}
Here 
\begin{equation}
D_k^{\#}(\cA)=  D (\cA ) + m_k  + b_k\cQ_{k}(-\cA)^T \cQ_{k}(\cA) 
\end{equation}
Under the conditions  on  $e_k | \pa \cA|$  this  operator is invertible
(about which more later)  and 
we define
\begin{equation}
\begin{split}
S_k(\cA) =& D_k^{\#}(\cA)^{-1}\\
\cH_k( \cA) =& 
\begin{cases}    
 b_k \  S_k(\cA)    \cQ_k( -\cA)^T  &   \textrm{ on } \Psi_k \\
 b_k \ S_k(\cA)^T  \cQ_k( \cA )^T  &   \textrm{ on } \bPsi_k \\
\end{cases}
 \\
D_k(\cA)  =&  b_k - b_k^2 \cQ_{k}(\cA)  S_k(\cA) \cQ_{k}(-\cA)^T \\
\end{split}
\end{equation}
Also   $\int[\dots] d \mu_{S_k(\cA)}(\psi)$  is the fermion Gaussian integral
with covariance $S_k(\cA)$.

To prove   (\ref{new}) 
make  the transformation   $\psi \to  \psi  +  \cH_k (\cA) \Psi_k$ and similarly for 
$\bpsi$.  This 
diagonalizes the quadratic form  and gives   
$ (\bPsi_k, D_k(\cA) \Psi_k)  + (\bpsi,  D_k^\#(\cA)  \psi)$,
whence the result with 
\begin{equation}
Z_k (\cA) =   \cM_{k,b_k}^{-1}   
\int e^{ -   ( \bpsi,D_k^\# (\cA)\psi)}  \psi 
\end{equation}

All these objects are gauge covariant.   In particular for a scalar $\la$  on $\bbT^{-k}_{N+M-k}$
\begin{equation}
S_k(\cA + d \la)   =   e^{-ie_k \la} S_k(\cA)  e^{ie_k \la}
\end{equation}

\subsubsection{}
We specifically  consider the generating function 
\begin{equation} 
\begin{split}
&\Om_k(\cA, \Psi_k,\eta)  =
 \int d \psi  \cM_{k,b_k}^{-1}
 \exp \left(
 - b_k   |\Psi_{k}- \cQ_{k}( \cA) \psi|^2  
\right)  
\exp \left( - (\bpsi,(D_{e_k}(\cA)+m_k) \psi) 
\right)  e^{  (\bar  \eta,  \psi) + (\bar \psi,  \eta) }\\
\end{split}
\end{equation}
where $\eta,\bar   \eta$ are elements of an auxiliary Grassmann algebra indexed by 
 $\bbT^{-k}_{N+M-k}$.
Using     (\ref{new})   and $ \int  \exp( (\bar  \eta,  \psi) + (\bar \psi,  \eta) )
d \mu_{S_k(\cA)} (\psi)=   \exp (   (\bar \eta, S_k(\cA) \eta))$
 this can be evaluated as 
 \begin{equation} 
\begin{split}
&\Om_k(\cA, \Psi_k,\eta)  =  \cZ_k(\cA)
\exp \left(  -  (\bPsi_k,  D_k(\cA)  \Psi_k)  
+ (\bar  \eta,  \cH_k (\cA) \Psi_k) + ( \cH_k (\cA) \bPsi_k,  \eta) +
 (\bar \eta,  S_k(\cA)  \eta) \right) \\
\end{split}
\end{equation}

We can also take it one step at a time.  By    (\ref{newt}) we have  for $\cA$ on 
   $\bbT^{-k-1}_{N+M-k-1}$
\begin{equation} 
\Om_{k+1}(\cA, \Psi_{k+1},\eta)  = \int \cM_{k+1,b}^{-1} \exp \left(
-  \frac{b}{L}  |\Psi_{k+1,L}- Q_{e_k}(  \tilde \cQ_k \cA_L) \Psi_{k}|^2   \right)   \Om_k(\cA_L,\Psi_k,  (
\si_{L^{-1}})^T
\eta)
\end{equation}
Put in the expression for   $\Om_k$ and 
 evaluate this by introducing 
\begin{equation}
\begin{split}
\Ga_k(\cA) =&  ( D_k(\cA)   + \frac{ b}{L}  Q_{e_k}(-\tilde \cQ_k \cA)^T Q_{e_k}(\tilde \cQ_k \cA) )^{-1}\\
H_k( \cA) =& 
\begin{cases}    
\frac{b}{L}  \Ga_k(\cA) Q_{e_k}(-\tilde \cQ_k \cA)^T  &   \textrm{ on } \Psi_k \\
\frac{b}{L}  \Ga_k(\cA)^T Q_{e_k}(\tilde \cQ_k \cA)^T  &   \textrm{ on } \bPsi_k \\
\end{cases}
 \\
\end{split}
\end{equation}
and making the transformation  $\Psi_k \to   \Psi_k +   H_{k}(\cA_L) \Psi_{k+1,L}$ and similarly 
for   $\bPsi_k$.
We get an alternative expression for $\Om_{k+1}$ and by comparing 
we find 
\begin{equation}  \label{nifty}
\begin{split}
S_{k+1}(\cA)   = & \si_L^{-1}   (S_{k}(\cA_L)  +  \cH_k(\cA_L)  \Ga_k(\cA_L) \cH_k(\cA_L)^T  )
(\si_L^{-1})^T
\\
\cH_{k+1}(\cA)   = &  \si_L^{-1}   \cH_k(\cA_L)  H_k(\cA_L) \si_L  \\
D_{k+1}(\cA)   =& \si_L^T   \left(\frac{b}{L} 
  -  \frac{b^2}{L^2}  Q_{e_k}(\tilde \cQ_k \cA_L) \Ga_k(\cA_L) Q_{e_k}(-\tilde \cQ_k \cA_L)^T\right)
\si_L   \\
\end{split}
\end{equation}
(The convention is that  $S_0(\cA)=0,  \cH_0(\cA)= I$, and  $D_0(\cA) = D_{e_0}(\cA)
+m_0 $.)  We also find 
that  
\begin{equation}  \label{iffy}
 Z_{k+1} (\cA) =
Z_k(\cA_L) \ \cM_{k+1,b}^{-1} \int  e^{- (\bPsi_k, \Ga_k(\cA_L)^{-1} \Psi_k)} d \Psi_k
\end{equation}

\subsubsection{}

For the kernels we can rewrite (\ref{nifty}) as  
\begin{equation}
S_{k+1} (\cA,x,x')  = 
 L^2  (S_k(\cA_L,Lx,Lx')  +  ( \cH_k(\cA_L)  \Ga_k(\cA_L) \cH_k(\cA_L)^T)(Lx,Lx') )
\end{equation}
Here we have used   $(\si_L^{-1})^T = L^{-1}\si_{L}$  for fermions.
Iterating this yields   
\begin{equation}  \label{iteratef}
S_k(\cA,x,x')  =
\sum_{j=0}^{k-1}L^{2(k-j)} \tilde \Ga_j(\cA_{L^{k-j}}; L^{k-j}x,
L^{k-j}x') 
\end{equation}
where 
 \begin{equation}
\tilde \Ga_j(\cA)  = \cH_j(\cA)\Ga_j(\cA) \cH_j(\cA)^T
\end{equation}
provided all the operators exist.

Balaban, O'Carroll, and Schor show    for  $\cA$ on  $\bbT^{-k}_{N+M-k}$ that  if
$e_k | \pa  \cA|$ is sufficiently small then  $S_k(\cA), \cH_k(\cA), \Ga_k(\cA)$
all exist and   
\begin{equation}
\begin{split}
|\cH_k(\cA, x,y)|  \leq & \cO(1) \exp( - \cO(1) d(x,y) )\\
|\Ga_k(\cA,y,y')|  \leq &  \cO(1) \exp( - \cO(1) d(y,y')) \\
|\tilde  \Ga_k(\cA,x,x')|  \leq &  \cO(1) \exp( - \cO(1) d(x,x')) \\
 \end{split}
\end{equation}
For  $\cA=0$ this can be found in  \cite{BOS89}.  For    $\cA \neq 0 $ 
it is a special case of results in  \cite{BOS91} and    section \ref{three}. 
 If   $\cA$ on  $\bbT^{-k}_{N+M-k}$ has  
$e_k |\pa  \cA|$ sufficiently small, then  $e_j |\pa  \cA_{L^{k-j}}|$ on  $\bbT^{-j}_{N+M-j}$ is even
smaller by  a factor of  $L^{-3(k-j)/2} $ so we can use   (\ref{iteratef}) to  obtain the bound
\begin{equation}  \label{duo}
| S_k(\cA,x,x')| \leq 
 \left\{ \begin{array}{rl}
\cO(1) d(x,x')^{-2} e^{- \cO(1) d(x, x'  )}  &  x\neq x' \\
\cO(L^{2k} )  & x=x'\\  
\end{array}  \right.  
\end{equation}
The function  $S_k(\cA, x,x')$ is our basic fermion  propagator
with background field $\cA$.  The estimate shows  the characteristic
short distant singularity  $d(x,x')^{-2}$.

\newpage

\subsection{global flow}

Now we combine the steps for bosons and fermions and make a first pass
at the global flow.  Our goal is not yet complete control.  We just 
want to introduce some  notation and establish 
some identities. 

We repeatedly apply the basic transformation (\ref{firstrep})
with fermion averaging operator $ Q_{e_k}(  A'_k) $ taken to be   $Q_{e_k}( \tilde \cQ_k  \cA_k)$   with  
$\cA_k = \cH_k A_k$. This choice  of  $\cA_k$ is made to match the background field at the $k^{th}$ step.
Then we have   
\begin{equation}
\begin{split}
 & \rho_{k}(\Psi_k, A_k)  = \int\prod_{j=0}^{k-1}dA_j \cN_{j+1,a}^{-1}   \exp \left(
- \frac12 \frac{a}{L^2}  |A_{j+1,L}- Q A_{j}|^2  \right) 
 \exp \left(- \frac12  ( A_0,(-\De+ \mu^2_0) A_0) \right)\\
&  \prod_{j=0}^{k-1} d \Psi_j \cM_{j+1,b} ^{-1} \exp \left(
 - \frac{b}{L}   |\Psi_{j+1,L}- Q_{e_j}(\tilde \cQ_j  \cA_j) \Psi_j|^2  
\right)   \exp \left( - (\bPsi_0,(D_{e_0}(\cA_0)+m_0) \Psi_0)
\right) \\
\end{split}
\end{equation}
Integrating out the intermediate fermions by (\ref{old}) we have 
\begin{equation}
\begin{split}
 & \rho_{k}(\Psi_k, A_k)  = \int\prod_{j=0}^{k-1}dA_j \cN_{j+1,a}^{-1}   \exp \left(
- \frac12 \frac{a}{L^2}  |A_{j+1,L}- Q A_{j}|^2  \right) 
 \exp \left(- \frac12  ( A_0,(-\De+ \mu^2_0) A_0) \right)\\
& \int d \psi  \cM_{k,b_k} ^{-1}  \exp \left(
 - b_k   |\Psi_{k}- \cQ_{k}(\cA_{k-1,L^{-1}}, \dots , \cA_{0,L^{-k}}) \psi|^2  
\right) \exp \left( - (\bpsi,(D_{e_k}(\cA_{0,L^{-k}})+m_k) \psi) 
\right)  \\
\end{split}
\end{equation}
We cannot now integrate out the intermediate boson fields, but 
 we   can    successively apply the transformations $A_j \to   A_j +   H_{j} A_{j+1}$
for   $j=0, \dots , k-1$  and use the identities   (\ref{iterate}).  Under the transformation
on  $A_j$ we have    $\cA_j  \to 
\cA_j +\cA_{j+1,L}$. Under all subsequent   transformations 
we have   
$\cA_j  \to   \sum_{i=j}^{k}   \cA_{i,L^{i-j}}$ 
and thus  $(\cA_j)_{L^{-(k-j)}} \to  \cA_{j,k}^*$ defined by 
\begin{equation}
   \cA_{j,k}^* =  \sum_{i=j}^{k}   (\cA_i)_{L^{-(k-i)}} 
\end{equation}
Thus we obtain
\begin{equation}  \label{star1}
 \rho_{k}(\Psi_k, A_k)  =  Z_k \exp \left( -    \frac12 (A_k,\De_{k}  A_k) 
 \right) \  \rho^{\star}_{k}(\Psi_k, \cA_k) 
\end{equation}
where 
\begin{equation}   \label{star2}
\begin{split}
  \rho^{\star}_{k}(\Psi_k, \cA_k)  = &   \int    \cM_{k,b_k}^{-1}
\exp \left(-b_k | \Psi_{k} -\cQ_k(\cA^*_{k-1,k},     \dots, \cA^*_{0,k})\psi|^2
\right) \\ &
 \exp \left(  -   (\bpsi ,( D_{e_k} ( \cA_{0,k}^*     )
+ m_k) \psi )  \right)  d\psi  \prod_{j=0}^{k-1}   d \mu_{C_j}(A_j) \\
\end{split}
\end{equation}
Next  for  $a_i$ on  $\bbT^{-k}_{N+M-k}$  define   
$a_{j,k}^* =  \sum_{i=j}^{k}  a_i$  and a potential  $\cV_k(\Psi_k, \psi, a_k,  \cdots , a_{0})$
by 
\begin{equation}
\begin{split}
& b_k | \Psi_{k} -\cQ_k(a^*_{k-1,k},     \dots, a^*_{0,k})\psi|^2 
+  (\bpsi , D_{e_k}( a^*_{0,k} )\psi)\\
 = &  
 b_k | \Psi_{k} -\cQ_k(a_k)\psi|^2 
+  (\bpsi , D_{e_k}( a_{k} )\psi)
 +\cV_k(\Psi_k, \psi, a_k,  \cdots , a_{0})\\
\end{split}
\end{equation}
Note that   $\cV_k(\Psi_k, \psi, a_k, 0, \cdots , 0)=0$. 
If we evaluate at   $a_i =  (\cA_i)_{L^{-(k-i)}}$ for  $i=0, \dots , k$ we get the 
argument of the exponential in  (\ref{star2}).  Thus   
\begin{equation}  
\begin{split}
  \rho^{\star}_{k}(\Psi_k, \cA_k)  = &   \int  \cM_{k,b_k}^{-1}
\exp \left(-b_k | \Psi_{k} -\cQ_k(\cA_k)\psi|^2
-  (\bpsi , D_{e_k} (\cA_k) \psi) \right)  \\
&   \exp \left(-\cV_k(\Psi_k, \psi, \cA_k, \cA_{k-1,L^{-1}}, \cdots , \cA_{0, L^{-k}}) \right)   d\psi 
\prod_{j=0}^{k-1} d \mu_{C_j}(A_j) \\
\end{split}
\end{equation}
Now if we assume $e_k|\pa \cA_k| $  is sufficiently small
we can evaluate  the fermion integral  by (\ref{new}). (This
assumption would not be suitable for iteration.)
Defining    $\psi_k(\cA_k) \equiv \cH_k(\cA_k)\Psi_k$ this yields 
\begin{equation}  \label{E}
 \rho^{\star}_k(\Psi_k, \cA_k) 
=  Z_k(\cA_k)\exp(- (\bPsi_k, D_k(\cA_k) \Psi_k) )\rho^{\bullet}_{k}(\Psi_k,  \psi_k(\cA_k),  \cA_k) 
\end{equation}
where  for   $\psi, \cA$ on  $\bbT^{-k}_{N+M-k}$
\begin{equation}
\begin{split}
&\rho^{\bullet}_{k}(\Psi_k,  \psi, \cA)
 = 
\int  \exp \left( -\cV_k(\Psi_k, \psi + \psi',\cA, \cA_{k-1,L^{-1}}, \cdots , \cA_{0, L^{-k}}) )
  \right)  d
\mu_{S_k(\cA) }(\psi')
\prod_{j=0}^{k-1}   d\mu_{C_j}(A_j) \\
\end{split}
\end{equation}
 Note that  $\rho^{\bullet}_{k}(\Psi_k,  \psi_k(\cA_k),  \cA_k)$ is actually just a function 
of  $\Psi_k,A_k$  but we find it convenient to keep track of the 
dependence in the variables   $\Psi_k,  \psi_k(\cA_k),\cA_k$.   This will be especially useful 
for local versions later on. 

Putting everything together we have  
\begin{equation} 
 \rho_{k}(\Psi_k, A_k)  =  Z_k Z_k(\cA_k)\exp \left( -    \frac12 (A_k,\De_{k}  A_k) 
- (\bPsi_k, D_k(\cA_k) \Psi_k) \right)\rho^{\bullet}_{k}(\Psi_k, \psi_k(\cA_k),  \cA_k) 
\end{equation}
This separates   off a kinematic part, 
and we now proceed to the study  $\rho^{\bullet}_{k}(\Psi_k,  \psi_k(\cA_k)
\cA_k) $ which is the interaction part.

\subsection{perturbation theory}

For perturbation theory we introduce a parameter  $t$
and define   instead of  (\ref{star2})
 \begin{equation}   \label{star3}
\begin{split}
  \rho^{\star}_{k}(t,\Psi_k, \cA_k)  = &   \int    \cM_{k,b_k}^{-1}
\exp \left(-b_k | \Psi_{k} -\cQ_k(\cA^*_{k-1,k}(t),     \dots, \cA^*_{0,k}(t))\psi|^2
\right) \\ &
 \exp \left(  -   (\bpsi ,( D_{e_k} ( \cA_{0,k}^*(t)     )
+ m_k) \psi )  \right)  d\psi  \prod_{j=0}^{k-1}   d \mu_{C_j}(A_j) \\
\end{split}
\end{equation}
where
\begin{equation}
   \cA_{j,k}^*(t) =\cA_k +t \sum_{i=j}^{k-1}   (\cA_i)_{L^{-(k-i)}} 
\end{equation}
These reduce to the previous quantities at  $t=1$.
We continue to  assume  $e_k |\pa \cA_k| $ is sufficiently small,
repeat the steps in the last section, and   find 
\begin{equation}  \label{F}
 \rho^{\star}_k(t,\Psi_k, \cA_k) 
=  Z_k(\cA_k)\exp(- (\bPsi_k, D_k(\cA_k) \Psi_k) )\rho^{\bullet}_{k}(t,\Psi_k,  \psi_k(\cA_k),  \cA_k) 
\end{equation}
where 
\begin{equation}  \label{autumn2}
\begin{split}
&\rho^{\bullet}_{k}(t,\Psi_k,  \psi, \cA)
 = 
\int  \exp \left( -\cV_k(\Psi_k, \psi + \psi',\cA, t\cA_{k-1,L^{-1}}, \cdots , t\cA_{0, L^{-k}}) )
  \right)  d
\mu_{S_k(\cA) }(\psi')
\prod_{j=0}^{k-1}   d\mu_{C_j}(A_j) \\
\end{split}
\end{equation}

We study  $\rho^{\bullet}_{k}(t)=\rho^{\bullet}_{k}(t,\Psi_k,  \psi, \cA)$ at  $t=1$
 by expanding it around  $t=0$: 
$\rho^{\bullet}_{k}(1) =\rho^{\bullet}_{k}(0)  +(\rho^{\bullet}_{k})'(0) +  (\rho^{\bullet}_{k})''(0)/2 +
\dots
$ Write (\ref{autumn2}) as      $\rho^{\bullet}_{k}(t) =< \exp(-\cV_{k}(t)>$ where  $<\cdots>$ indicates
the Gaussian integrals. 
Then $\rho^{\bullet}_{k}(0)=< \exp(-\cV_{k}(0)>  =1$ since  $\cV_{k}(0) =0$
and    $(\rho^{\bullet})'_{k}(0) =< -\cV'_{k}(0)> =0$ since each  term in $\cV'(0)$ is odd in
some boson field.  Thus the first non-trivial term  
$(\rho^{\bullet}_{k})''(0)/2$   which we study further.   (This is also  
the first non-zero term an expansion of  the effective potential $ \log(\rho^{\bullet}_{k}(1))$
 around  $t=0$. )

Accordingly we define
\begin{equation}  
  P_k( \Psi_k, \psi,  \cA) \equiv   \frac12   ( \rho^{\bullet}_{k})''(0) 
=   \frac12 <\cV'_k(0) ^2> -  \frac12 <\cV''(0)>  
\end{equation}
The $t$ derivatives can be evaluated as   $a_j$ derivatives and 
we also use  
\begin{equation}
\int   \cA_{j,L^{-(k-j)}}(z)\cA_{j,L^{-(k-j)}}(w)  d\mu_{C_j}(A_j)
=   L^{k-j}\tilde  C_j( L^{k-j}z, L^{k-j}w)  \equiv  \tilde  C_{j,L^{-(k-j)}}( z, w)  
\end{equation}
Then the 
boson fluctuation integrals can be evaluated as
\begin{equation}   \label{bosonf}
\begin{split}  
&   P_k(\Psi_k, \psi,  \cA) \\
=
&\frac12 \sum_{j=0}^{k-1}  \int    \frac{\de \cV_k }{\de a_{j}(z)  } 
( \Psi_k, \psi+ \psi', \cA,0)
  \frac{\de\cV_k }{\de a_{j}(w)   } 
( \Psi_k, \psi+ \psi', \cA,0)
 \tilde  C_{j,L^{-(k-j)}}(z,w)  d\mu_{S_k(\cA)}(\psi')\\
&
- \frac12 \sum_{j=0}^{k-1}  \int    \frac{\de^2 \cV_k }{{\de a_{j}(z)\de a_{j}(w)} } 
( \Psi_k, \psi+ \psi', \cA,0) \tilde  C_{j,L^{-(k-j)}}( z,w)  d\mu_{S_k(\cA)}(\psi')\\
\end{split}
\end{equation}
Now do  the fermion Gaussian integral, drop the no fermion part from $  P_k$ and call the rest  $  P^+_k $,
and  we find
\begin{equation}    \label{pexprime}
\begin{split}   P^+_k(\Psi_k, \psi,  \cA)
= &
\frac12 \sum_{j=0}^{k-1}  \int \cJ_{kj}(z)
   \tilde  C_{j,L^{-(k-j)}}(z,w)   \cJ_{kj}(w)
- \int   \cJ_{kj} (z,w)
  \tilde  C_{j,L^{-(k-j)}}(z,w) \\
 - &  \sum_{j=0}^{k-1}  \int    \bar  \cK_{kj}(x,z)
   S_k (\cA;x,y)   \cK_{kj}(y,w)\ \tilde  C_{j,L^{-(k-j)}}(z,w) \\
\end{split}
\end{equation}
where  the vertices are
\begin{equation}   \label{vertex}
\begin{split}
\cJ_{kj}(z) 
=\frac{\pa  \cV_k }{  \pa  a_{j}(z)    } ( \Psi_k, \psi,\cA,0) 
& 
\ \ \ \ \ \ \ \ \ \ 
\cJ_{kj}(z,w) 
=\frac{\pa ^2 \cV_k }{  \pa  a_{j}(z)\pa  a_{j}(w)  } ( \Psi_k, \psi,\cA,0)
\\
\bar \cK_{kj}(x,z) 
=\frac{\pa ^2 \cV_k }{\pa  \psi(x) \pa  a_{j}(z)   } ( \Psi_k,\psi,\cA,0)
&
  \ \ \ \ \ \ \ \ \ \ \ \ \
\cK_{kj}(y,z) 
=\frac{\pa ^2 \cV_k }{\pa  \bpsi(y) \pa  a_{j}(z)  } ( \Psi_k,\psi,\cA,0) 
\\
\end{split}
\end{equation}
The terms in  (\ref{pexprime})    can be labeled by Feynman diagrams, see paper  I.

We study   
 $\cJ_{kj, \mu }(z) 
=\pa  \cV_k /  \pa  a_{j, \mu }(z) $ further.  This has two parts
$\cJ_{kj, \mu }(z)  = \cJ^D_{kj, \mu }(z)   + \cJ^Q_{kj, \mu }(z).$
  The classical part   is 
\begin{equation}
\begin{split}
\cJ^{D}_{kj, \mu }(z)
=& \frac{\pa} { \pa a_{j,\mu}(z)}
 (\bar \psi, D_{e_k}(\cA  + \sum_{j=0}^{k-1}a_j))\psi)|_{a_j=0} 
=
 \frac{\pa} { \pa  \cA_{\mu}(z)} (\bar \psi,
D_{e_k}(\cA) \psi) \\
 = & ie_k
\ \bpsi(z) (\frac{r + \ga_{\mu}}{2}) \exp (ie_kL^{-k} \cA_{\mu}(z)) \psi(z+
L^{-k} e_{\mu})\\
 - &
 ie_k\ \bpsi(z + L^{-k} e_{\mu}) (\frac{r - \ga_{\mu}}{2}) \exp (-ie_kL^{-k}
\cA_{\mu}(z)) \psi(z)\\
\end{split}
\end{equation}
 The term is independent of  $j$.
 If this were the only 
contribution we could resum and get the full boson propagator
$\sum_{j=0}^{k-1}  \tilde  C_{j,L^{-(k-j)}}  =G_k$  as in  (\ref{decompb}).
Note that as  $k=N \to \infty$ it approaches the usual continuum current
$ie \bpsi(z) \gamma_{\mu} \psi(z)$.

The other part is an artifact of our renormalization group procedure.  It is 
\begin{equation}
\begin{split}
\cJ^{Q}_{kj, \mu }(z) =  & \frac{\pa} { \pa a_{j,\mu}(z)}    
  b_k | \Psi_{k} -\cQ_k(\cA +a^*_{k-1,k-1},    
\dots,\cA+  a^*_{0,k-1})\psi|^2 |_{a_{k-1}=0, \dots , a_0
=0}\\ =  & \frac{\pa} { \pa a_{j,\mu}(z)}    
  b_k | \Psi_{k} -\cQ_k(\cA, \dots, \cA,\cA+a_j, \dots,   \cA+a_j) \psi|^2 |_{a_j=0}\\
\end{split}
\end{equation}
where there are $j+1$ entries  with   $\cA+a_j$.

Similar expressions hold for   the other vertices in  (\ref{vertex}).

The expression  $P_k^+$ has no ultraviolet divergences, i.e. it is bounded as
$k \to \infty$.  This is true in this global version even without the counterterms.
In paper III we prove it for a local version, and in that case   the counterterms are 
needed.

\subsection{single steps}  \label{ss}

We want to get an identity relating  $P^+_k$ and $P^+_{k+1}$, and this  means 
 investigating   how the densities change under a single RG transformation.

Start with the identity
\begin{equation}  \label{winter}
\begin{split}
& \rho^{\star}_{k+1}(t,\Psi_{k+1}, \cA_{k+1})=
\int    d \Psi_k \ d\mu_{C_k}(A_k)\ \cM_{k+1,b}^{-1}  \\
&
\exp \left(-   \frac{b}{L}   | \Psi_{k+1,L} -
Q_{e_k}(\tilde \cQ_k ( \cA_{k+1,L} +
t\cA_k))\Psi_k|^2
\right) \rho^{\star}_{k}(t,\Psi_k,  \cA_{k+1,L} +t \cA_k)\\
\end{split}
\end{equation}
This can be proved by inserting the definition of $ \rho^{\star}_{k}(t)$ on
the right and using    (\ref{newt}).  We also use that replacing    $\cA_k$ by   
$\cA_{k+1,L} + t \cA_k$ has the effect of replacing   $\cA_{j,k}^*(t) $ by  $ [\cA_{j,k+1}^*(t) ]_L$.

We would like to  insert into this the expression (\ref{E}) giving  $\rho^{\star}$ in terms of a further 
reduced density  $\rho^{\bullet}$, and get a recursion relation  for   $\rho^{\bullet}$.
 However this expression for  $ \rho^{\star}_{k}$  is 
only valid with restrictions on  $\cA_k$ which we cannot assume
in the integral. For the moment we proceed  \textit{formally}.

Inserting (\ref{E}) into    (\ref{winter})  we have 
\begin{equation}
\begin{split}
& \rho^{\star}_{k+1}(t,\Psi_{k+1}, \cA_{k+1}) 
= 
\int   d\Psi_k\ d \mu_{C_k}(A_k) \cM_{k+1,b}^{-1} 
\exp \left(-   \frac{b}{L}   | \Psi -Q_{e_k}(\tilde \cQ_k(\cA+ t\cA_k))\Psi_k|^2
\right)  \\
&
\exp \left(
- (\bPsi_k, D_k(\cA+t \cA_k ) \Psi_k)\right)  \cZ_k(\cA +t \cA_k) 
\rho^{\bullet}_{k}(t,\Psi_k,\psi_k( \cA +t \cA_k),  \cA +t \cA_k)
|_{\stackrel{\Psi=\Psi_{k+1,L}}{ \cA = 
\cA_{k+1,L}}}\\
\end{split}
\end{equation}
Now  define  $V_k,U_k, \de \cH_k$ by 
\begin{equation}  \label{secondv}
\begin{split}
V_k(\Psi, \Psi_k,  \cA, \cA_k) =&    \frac{b}{L}   | \Psi -Q_{e_k}(\tilde \cQ_k(\cA+ \cA_k))\Psi_k|^2
+(\bPsi_k, D_k(\cA+ \cA_k ) \Psi_k)
-  \{\cA_k=0\} \\
 \cZ_k(\cA+ \cA_k)  =&  \cZ_k(\cA)   \exp (-U_k(\cA , \cA_k))\\
\de \cH_k(\cA, \cA_k) = & \cH_k(\cA + \cA_k) - \cH_k(\cA )\\
\end{split}
\end{equation}
We obtain
\begin{equation}
\begin{split}
 & \rho^{\star}_{k+1}(t,\Psi_{k+1}, \cA_{k+1}) \\
= & \cZ_k(\cA )  
\int  d\Psi_k\ d \mu_{C_k}(A_k)  \cM_{k+1,b}^{-1} 
\exp \left(-  \frac{b}{L}    | \Psi -Q_{e_k}(\tilde \cQ_k\cA )\Psi_k|^2
- (\bPsi_k, D_k(\cA ) \Psi_k)\right)\\
&
\exp\left( -V_k(\Psi, \Psi_k,  \cA,t \cA_k) -U_k(\cA ,t \cA_k)\right)
\rho^{\bullet}_{k}(t,\Psi_k,\psi_k(\cA) +\de \cH_k(\cA, t\cA_k)\Psi_k, \cA + t\cA_k)
|_{\stackrel{\Psi=\Psi_{k+1,L}}{ \cA = \cA_{k+1,L}}}\\
\end{split}
\end{equation}
Now make the translation  $\Psi_k \to \Psi_k +  H_{k}(\cA)\Psi \equiv  \Psi_k + \Psi(\cA) $ 
This means that   $\psi_k(\cA_{k+1,L})  \to   [ \psi_{k+1}(\cA_{k+1})]_L  +\psi_k(\cA_{k+1,L})  $.
Using      (\ref{nifty}) and (\ref{iffy}) and identifying a Gaussian measure we find 
\begin{equation}  \label{spring}
\begin{split}
 & 
\rho^{\bullet}_{k+1}(t,\Psi_{k+1},\psi_{k+1}(\cA_{k+1}), \cA_{k+1}) \\
=&
\int  d \mu_{\Ga_k(\cA)}(\Psi_k)\ d \mu_{C_k}(A_k)    
\exp  ( -V_k(\Psi,\Psi(\cA) + \Psi_k, \cA, t\cA_k)
 -U_k(\cA ,t\cA_k))   
\\
 &
 \rho^{\bullet}_k(t,\Psi(\cA)  + \Psi_k, [ \psi_{k+1}(\cA_{k+1})]_L  +\psi_k(\cA)+
\de \cH_k(\cA,t \cA_k)(\Psi(\cA)  + \Psi_k),\ 
\cA+t\cA_k  )
 |_{\stackrel{\Psi=\Psi_{k+1,L}}{ \cA = \cA_{k+1,L}}}
 \\
\end{split}
\end{equation}
This is our recursion relation for   $\rho^{\bullet}_k(t)$.

Now take two derivatives of 
the last equation at $t=0$ and obtain
\begin{equation} 
\begin{split}
&  P_{k+1}(\Psi_{k+1},\psi_{k+1}(\cA_{k+1}),\cA_{k+1}) 
=
\int P_{k}( \Psi(\cA) +\Psi_k,[ \psi_{k+1}(\cA_{k+1})]_L +  \psi_k(\cA),\ \cA)
 d\mu_{\Ga_k( \cA)}(\Psi_k) \\
+&  \frac12
\int  \frac{ \pa V_k}{ \pa \cA_{k}(z)} (\Psi, \Psi(\cA)  +\Psi_k, \cA,0)
\tilde C_k(z,w) 
 \frac{ \pa V_k}{ \pa \cA_{k}(w)} (\Psi, \Psi(\cA)  +\Psi_k, \cA,0) 
dz dw\ d\mu_{\Ga_k( \cA)}(\Psi_k) \\
-& \frac12  \int  \frac{ \pa^2 V_k}{ \pa \cA_{k}(z) \pa \cA_{k}(w)} ( \Psi, \Psi(\cA)  +\Psi_k, \cA,0)
\tilde C_k(z,w)\  dz\ d\mu_{\Ga_k( \cA)}(\Psi_k)
 |_{\stackrel{\Psi=\Psi_{k+1,L}}{ \cA = \cA_{k+1,L}}}  +   \dots \\
\end{split}
\end{equation}
where  $\dots$ indicates no-fermion terms.
Next we carry out  the fermion integrals and drop the no fermion terms  
to get
the expression
 \begin{equation}  \label{basic}
\begin{split}
&  P^+_{k+1}(\Psi_{k+1},\psi_{k+1}(\cA_{k+1}),\cA_{k+1}) \\
= &  \left[  P^+_k( \Psi(\cA)  ,[ \psi_{k+1}(\cA_{k+1})]_L,\ \cA)
-  (\De ^{\Ga}_{ \Psi_k}P_{k} )^+  ( \Psi(\cA)  ,[ \psi_{k+1}(\cA_{k+1})]_L,\ \cA) \right.
 \\  
+  & \frac12   \int_{z,w}  J_{k}(z)
\tilde   C_k(z,w)  J_{k}(w)-   J_{k} (z,w)\tilde  C_k(z,w) \\
 - & \left.    \int_{z,w} \sum_{x,y}   \bar  K_{k}(z,x)
\Ga_k ( \cA;x,y)   K_{k}(w,y) \tilde  C_k(z,w)
 \right]_{\stackrel{\Psi =\Psi_{k+1,L}}{ \cA =  \cA_{k+1,L}}} \\
\end{split}
\end{equation}
where   the single step vertices are  
\begin{equation} \label{vertex2}
\begin{split}
J_{k}( z) 
=   \frac{\pa V_k} {\pa \cA_{k}(z) } 
(\Psi, \Psi(\cA) , \cA,0) & \ \ \ \ \ \ \ \ J_{k}( z,w) 
=   \frac{\pa^2 V_k} {\pa \cA_{k}(z)\pa \cA_{k}(w) } (\Psi,  \Psi(\cA)  , \cA,0)
\\
\bar  K_k (z,x) 
=   \frac{\pa^2 V_{k} }{\pa  \Psi_k(x)\pa \cA_{k,\mu}(z) } 
 (  \Psi, \Psi(\cA) , \cA,0)
  & \ \ \ \ \ \ \ \ \ 
  K_k(z,x) 
=    \frac{\pa^2 V_{k} }{\pa  \bPsi_k(x)\pa \cA_{k,\mu}(z) } 
 (  \Psi, \Psi(\cA) , \cA,0)
\\
\end{split}
\end{equation}
and the  notation is  
\begin{equation}
\begin{split}
&(\De ^{\Ga}_{ \Psi_k}P_{k} )( \Psi(\cA)  ,[ \psi_{k+1}(\cA_{k+1})]_L,\ \cA) \\
=&
   \sum_{x,y}  \Ga_k(\cA,x,y) \left[  \frac{\pa^2}{ \pa   \Psi_k(x)  \pa  \bar   \Psi_k(y)
}    P_{k} ( \Psi(\cA)  +\Psi_k, [\psi_{k+1}(\cA_{k+1})]_L +  \cH_k(\cA)\Psi_k,\ \cA) \right]_{\Psi_k =0}\\
\end{split}
\end{equation}

The equation  (\ref{basic}) is  the basic identity we are after.  Although the derivation was
formal we show in  appendix 
\ref{B}  
 that  it is rigorously true provided $e_{k+1}| \pa \cA_{k+1}|$ is sufficiently small.
\bigskip

We also  need a variation of   (\ref{basic}) in which vertices are localized
in  a region  $\Theta  \subset  \bbT^{-k}_{N+M-k}$.    First define   $\rho_{k, \Theta}^{\star}(t)$
just as in (\ref{star3}) but now with  
 $\cA_{j,k}^*(t)$ replaced by    
\begin{equation}
   \cA_{j,k}^*(t, \Theta) =\cA_k +t  \chi_{\Theta}\sum_{i=j}^{k-1}   (\cA_i)_{L^{-(k-i)}} 
\end{equation}
where  $ \chi_{\Theta}$ is the characteristic function of $\Theta$.
Then 
$ \rho^{\star}_{k, \Theta}(t)
=  Z_k(\cA_k)\exp(- (\bPsi_k, D_k(\cA_k) \Psi_k) )\rho^{\bullet}_{k,\Theta}(t)$
where     
$\rho_{k, \Theta}^{\bullet}(t)$
defined as  in   (\ref{autumn2}) except that      
$t\cA_{j,L^{-(k-j)}}$  is replaced by   $t  \chi_{\Theta}\cA_{j,L^{-(k-j)}}$. 
Defining   $P_{k, \Theta} =  (\rho_{k, \Theta}^{\bullet})''(0)/2$
we find
\begin{equation}    \label{pexprime2}
\begin{split}   P^+_{k,\Theta}(\Psi_k, \psi,  \cA)
= &
\frac12 \sum_{j=0}^{k-1}  \int_{z,w \in \Theta} \cJ_{kj}(z)
   \tilde  C_{j,L^{-(k-j)}}(z,w)   \cJ_{kj}(w)
 -  \cJ_{kj} (z,w)
  \tilde  C_{j,L^{-(k-j)}}(z,w) \\
 - &  \sum_{j=0}^{k-1}  \int_{z,w \in \Theta}     \bar  \cK_{kj}(z,x)
   S_k (\cA;x,y)   \cK_{kj}(w,y)\ \tilde  C_{j,L^{-(k-j)}}(z,w) \\
\end{split}
\end{equation}

One can also proceed in single steps.  One shows for   $\Theta  \subset  \bbT^{-k-1}_{N+M-k-1}$  that 
$\rho_{k+1, \Theta}^{\bullet}(t)$ and  $\rho_{k, L\Theta}^{\bullet}(t)$ are  formally related by an
equation like   (\ref{spring}) except that under the integral sign  $ t\cA_k$ is everywhere replaced by 
 $t \chi_{\Theta} \cA_k$.
Taking two derivatives at  $t=0$ yields the identity
 \begin{equation}  \label{basic2}
\begin{split}
&  P^+_{k+1,\Theta}(\Psi_{k+1},\psi_{k+1}(\cA_{k+1}),\cA_{k+1}) \\
= &  \left[  P^+_{k,L \Theta}( \Psi(\cA)  ,[ \psi_{k+1}(\cA_{k+1})]_L,\ \cA)
-  (\De ^{\Ga}_{ \Psi_k}P_{k,L\Theta} )^+  ( \Psi(\cA)  ,[ \psi_{k+1}(\cA_{k+1})]_L,\ \cA) \right.
 \\  
+  & \frac12   \int_{z,w \in L\Theta}  J_{k}(z)
\tilde   C_k(z,w)  J_{k}(w)
 -   J_{k} (z,w)\tilde  C_k(z,w) \\
 - & \left.  \int_{z,w \in L\Theta}  \sum_{x,y}    \bar  K_{k}(z,x)
\Ga_k ( \cA;x,y)   K_{k}(w,y) \tilde  C_k(z,w)
 \right]_{\stackrel{\Psi =\Psi_{k+1,L}}{ \cA =  \cA_{k+1,L}}} \\
\end{split}
\end{equation}
Again the derivation is formal, but the result is rigorous by the argument of 
appendix \ref B.

\newpage

\section{Localized flow}  \label{two}

\subsection{blocking}

We  also want to  consider a version of our RG transformation  in which the averaging
is not done on the whole torus, but in a sequence of  successively smaller 
regions.  The treatment follows Balaban, O'Carroll, and Schor  \cite{BOS91}.
A difference is that they do not make the initial scaling to a unit lattice as in (\ref{rho0}).
This means they are working up from a finer to a coarser lattice, whereas we are working down from a
coarser to a finer lattice.

First we define some blocking and unblocking operations.  For $\Om \subset \bbT^1_n \subset   \bbT^0_n$ 
we defined  a blocked set  $B\Om  \subset  \bbT^0_n  $   by 
\begin{equation}
B \Om  =   \{x \in  \bbT^0_{n}:  d(x, \Om) < L/2 \}
\end{equation}
Then  $QA$ on $\Om$ depends on  $A$ on  $B\Om$,  written  $(QA)_{\Om} = Q(A_{B\Om})$.
A set  $\La  \subset   \bbT^0_n$ has the form $B \Om$  iff it
   is a union of  $L$-blocks in  $\bbT^0_n$ centered on points in 
$\bbT^1_n$

For   $\La  \subset   \bbT^0_n$ we also define an   unblocked  $ \La' \subset  \bbT^1_n $      by 
\begin{equation}
 \La'  = U \La  =   \La   \cap  \bbT^1_n
\end{equation}
We have always   $UB \Om = \Om$.  If $\La = B\Om$  then 
  $B U \La = \La$. 
 Note    that if  $\La \subset \bbT^0_n$ then  $BL\La  \subset \bbT^0_{n+1}$ and 
 $L^{-1}U \La   \subset \bbT^0_{n-1}$.  More generally we define for  $\La \subset \bbT^0_n$ 
\begin{equation}
\begin{split}
B_{\ell} \La   = & (BL)^{\ell} \La  \subset   \bbT^0_{n+ \ell} \\
U_{\ell} \La   = & (L^{-1}U)^{\ell} \La  \subset   \bbT^0_{n- \ell} \\
\end{split}
\end{equation}
We have   $U_{\ell}B_{\ell}\La = \La$.

Our regions will be a sequence of  the form 
\begin{equation}
\bla = (\La_0,  \dots,  \La_{k-1}) 
\end{equation}  
where    $\La_j   \subset  \bbT^0_{N+M-j}$  is a union of 
$LM_0 = L^{m_0+1}$ blocks centered
on   $\bbT^{m_0+1}_{N+M-j}$, $m_0 \geq 1$.
We assume the sets are decreasing in the sense that they satisfy   one of the equivalent
\begin{equation}
B_1\La_{j+1} \subset  \La_j  \ \  \ \ \ \ \ \ \   \La_j  \subset   U_1\La_{j-1}= L^{-1}  \La_{j-1}' 
\end{equation}
We also assume that  for some positive integer  $r$
\begin{equation}  \label{separation}
d(  ( L^{-1}  \La_{j-1}')^c ,   \La_j )  \geq r M_0
\end{equation}
whenever both subsets are non-empty.
This  insures that the corridor between successive regions is 
at least a few $M_0$ blocks wide.

 We  define  in  $\bbT^0_{N+M-i}$
\begin{equation}
\de \La_i  = L^{-1} \La_{i-1}' - \La_i  
\end{equation}
Then with the convention  $\La_k = \emptyset$ 
we have    the disjoint union
\begin{equation}   \label{dis}
\La_j =  \cup_{i=j+1}^k  B_{i-j}   \de \La_i  
\end{equation}

\subsection{bosons}

We want a generalization of the formula (\ref{combine}) for bosons in which the
averaging over  $A_j$ is only done in the region  $\La_j$.
The  starting point is  
 \begin{equation}    \label{form1}
\int\prod_{j=0}^{k-1}dA_{j,\La_j} \cN_{L^{-1}\La_j',a}^{-1}   \exp \left(
- \frac12 \frac{a}{L^2}  |A_{j+1,L}- Q A_{j}|^2_{\La_j'}  \right) 
 \exp \left(- \frac12  ( A_0,(-\De+ \mu^2_0) A_0) \right) F(A_0) 
\end{equation}
where    $\cN_{\La,a} =  (2\pi/a)^{3|\La|/2}$. 
Break this up by   (\ref{dis})  
\begin{equation}
\begin{split}
& \cN_{L^{-1}\La_j',a}^{-1}   \exp \left(
- \frac12 \frac{a}{L^2}  |A_{j+1,L}- Q A_{j}|^2_{\La_j'}  \right) dA_{j, \La_j} \\
= & \prod_{i=j+1}^{k} \cN_{L^{-1}(B_{i-j}\de  \La_i)',a}^{-1} 
 \exp \left(
- \frac12 \frac{a}{L^2}  |A_{j+1,L}- Q A_{j}|^2_{(B_{i-j}\de  \La_i)' }  \right) 
 dA_{j ,B_{i-j}\de  \La_i} \\
\end{split}
\end{equation}
Change the order of the products
$
\prod_{j=0}^{k-1}  \prod_{i=j+1}^k  =  \prod_{i=1}^k
  \prod_{j=0}^{i-1}
$
 and evaluate the integral over   $\de \La_i$ by     
\begin{equation}
\begin{split}
&\int   \prod_{j=0}^{i-1}  dA_{j, B_{i-j}\de  \La_i} 
\cN_{L^{-1}(B_{i-j}\de  \La_i)',a}^{-1} 
 \exp \left(
- \frac12 \frac{a}{L^2}  |A_{j+1,L}- Q A_{j}|^2_{(B_{i-j}\de  \La_i)' } \right) [\dots ] \\
=&  
\int   dA_{0, B_i \de \La_i}   \cN_{\de \La_i,a_i}^{-1}   \exp \left(
- \frac{a_i}{2}  |A_{i}- \cQ_{i} A_{0,L^{-i}}|^2_{\de \La_i}  \right) [\dots ]\\
\end{split}
\end{equation}
This  follows from  (\ref{q2}) in 
the appendix, and then  (\ref{form1}) becomes 
\begin{equation}  \label{form1.5}
\int  \prod_{i=1}^{k}  
\cN_{\de \La_i,a_i}^{-1}   \exp \left(
- \frac{a_i}{2}   |A_i- \cQ_i A_{0,L^{-i}}|^2_{\de \La_i}  \right)
\exp \left(- \frac12  ( A_0,(-\De+ \mu^2_0) A_0) \right) F(A_0)  dA_{0, \La_0}  
\end{equation}
Next    scale $A_0=  \cA_{L^k}$ for    $\cA$ on  $\bbT^{-k}_{N+M-k}$   
and define  $\cQ_{i}^{(k)}=  \cQ_{i}
 \si_L^{k-i} $, which  is a $i$-fold averaging operator from
$\bbT^{-k}_{N+M-k}$ to  $\bbT^{0}_{N+M-i}$. 
Then we have that 
(\ref{form1.5}) is equal to 
\begin{equation}  \label{form2}
\int  \prod_{i=1}^{k}  
\cN_{\de \La_i,a_i}^{-1}   \exp \left(
- \frac{ a_i}{2}  |A_i- \cQ^{(k)}_i \cA|^2_{\de \La_i}  \right)
\left(- \frac12  ( \cA,(-\De+ \mu^2_k) \cA) \right) 
f(\cA)   d \cA_{L^{-k} \La_0}
\end{equation}
where $f(\cA)  = F( \cA_{L^k} )$.
In this formula the spectator  variables   $A_{0,\La^c}$ appear as   $\cA_{L^{-k} \La^c_0}   \equiv 
(A_{0,\La_0^c})_{L^{-k}}$.
\bigskip

We introduce  some  notation.
Let      
\begin{equation}
\begin{split}
\bfA'= &( A_{1,\de \La_1}, \cdots ,  A_{k, \de \La_{k}})\\
\cQ_{k, \bla} \cA  =  &(( \cQ^{(k)}_1  \cA)_{\de \La_1}, \cdots , (\cQ^{(k)}_k  \cA)_{\de \La_k})\\
\end{split}
\end{equation}
Note that since   $\La_k = \emptyset$ we have    $ A_{k, \de \La_{k}} =  A_{k, L^{-1} \La_{k-1}'}$.
These are   multiscale objects consisting of functions living on subsets $\de \La_i \subset
\bbT^0_{N+M-k}$.  A norm on such objects is given by 
\begin{equation}
| \bfA'|^2  =  \sum_{i=1}^k  | A_{i, \de \La_i} |^2
\end{equation}
We also define   $\bfa=(a_1, \dots, a_k)$ 
and     $  \cN_{\bla, \bfa} = \prod_{i=1}^{k}  \cN_{\de \La_i,a_i}^{-1} $. 
 Then (\ref{form2}) can be written    
\begin{equation}  \label{form3}
\int 
 \cN_{\bla, \bfa}^{-1} \exp \left(- \frac12   |\bfa^{1/2}(\bfA'- \cQ_{k, \bla} \cA ) |^2  \right)
\left(- \frac12  ( \cA,(-\De+ \mu^2_k) \cA) \right) 
f( \cA )   d \cA_{L^{-k} \La_0}
\end{equation}

To evaluate this integral  we again diagonalize the quadratic form,
this time separating the variables   $  (A_{0, \La^c},  \bfA')  $  or 
$ (\cA_{L^{-k} \La_0^c},  \bfA') $
from  $\cA_{L^{-k} \La_0}$.
Accordingly we introduce
\begin{equation}
\De_{k, \bla}^\# = -\De+ \mu^2_k  +   \cQ_{k,\bla}^T  \bfa   \cQ_{k,\bla} 
\end{equation}
and with $\bfA \equiv  (A_{0, \La^c},  \bfA') $
\begin{equation}
\begin{split}
G_{k,\bla } =&  [\De_{k, \bla}^{\#} ]^{-1}_{L^{-k}\La_0}\\
\cH_{k, \bla}  \bfA =&   G_{k,\bla } (\cQ_{k,\bla}^T \bfa  \bfA'  
  +[\De]_{\La,\La^c} \cA_{\La^c} ) |_{\La =   L^{-k} \La_0}
\\
( \bfA, \De_{k, \bla} \bfA) =   &
(\bfA ', \left[  \bfa -  \bfa \cQ_{k,\bla}  G_{k,\bla}\cQ_{k,\bla}^T \bfa\right] \bfA')
-   2(\cQ_{k,\bla}^T \bfa \bfA', 
  G_{k,\bla}  [\De]_{\La, \La^c}\cA_{\La^c}  )\\
&+( \cA_{\La^c}  \left[  ( -\De+\mu^2_k ) - 
 [\De_k]_{\La^c, \La}  G_{k,\bla}[\De]_{\La, \La^c}\right]\cA_{\La^c} )
|_{\La =   L^{-k}\La_0}\\
\end{split}
\end{equation}
Note that  $\cQ_{k,\bla} $ vanishes on functions on  $ L^{-k}\La^c_0$ and thus   
$\cQ_{k,\bla} ^T$ maps to functions on  $ L^{-k}\La_0$.  
Making the change of variables   $\cA_{L^{-k} \La_0} \to \cA_{ L^{-k} \La_0} + \cH_{k, \bla}  \bfA $ we find
that  (\ref{form3}) can be written   
\begin{equation}  \label{form4}
     Z_{k,\bla} \ \exp \left(- \frac 12 (\bfA,  \De_{k,\bla} \bfA)\right) 
\int   f( \cA+\cH_{k, \bla} \bfA )\  d\mu_{\st{G_{k,\bla}}} 
(\cA_{L^{-k} \La_0})
\end{equation}
where  
\begin{equation}
Z_{k,\bla} =\int 
 \cN_{\bla,\bfa}^{-1}\exp \left(- \frac12  ( \cA_{\La}, G_{k,\bla}^{-1}\cA_{\La}) \right) 
 d \cA_{\La}|_{\La= L^{-k} \La_0}
\end{equation}

Our final identity is then  (\ref{form1}) = (\ref{form4}).  We are particularly interested
in    $F=1$ in which case the identity reads
 \begin{equation}    \label{form5}
\begin{split}
& \int\prod_{j=0}^{k-1}dA_{j,\La_j} \cN_{L^{-1}\La_j',a}^{-1}   \exp \left(
- \frac12 \frac{a}{L^2}  |A_{j+1,L}- Q A_{j}|^2_{\La_j'}  \right) 
 \exp \left(- \frac12  ( A_0,(-\De+ \mu^2_0) A_0) \right)\\
&\ \ \ \ \ \ \ \ \ \  = Z_{k,\bla} \ \exp \left(- \frac 12 (\bfA,  \De_{k,\bla}
\bfA)\right)
\\
\\
\end{split}
\end{equation}

\subsection{fermions}

For fermions pick a fixed background  $\cA$ on  $\bbT^{-k}_{N+M-k}$ and consider 
integrals of the form 
\begin{equation}  \label{form 5.5}
\begin{split}
& \int   \prod_{j=0}^{k-1} d \Psi_{j, \La_j}   
 \cM_{L^{-1}\La_j',b}^{-1}  \exp \left(
  - \frac{b}{L} |\Psi_{j+1,L}- Q(\tilde \cQ_j\cA_{L^{k-j}} ) \Psi_{j}|^2_{\La_{j}}  \right) \\
&\ \ \ \ \ \ \ \ \exp \left( -
(\bPsi_0,(D_{e_0}( \cA_{L^k} )+m_0) \Psi_0)\right) F(\Psi_0) \\
\end{split}
\end{equation}
Doing the intermediate integrals as for bosons this can be written 
\begin{equation} \label{form6}
\begin{split}
 &\int \prod_{i=1}^{k} \cM_{\de \La_i,b_i}^{-1}   \exp \left(
  - b_i |\Psi_i- \cQ_i(\cA_{L^{k-i}}) \Psi_{0,L^{-i}}|^2_{\de \La_i}  \right) \\
&\ \ \ \ \ \ \ \exp \left( -
(\bPsi_0,(D_{e_0}(\cA_{L^k} )+m_0 )\Psi_0)\right)  F(\Psi_0)  d \Psi_{0,\La_0} \\
\end{split}
\end{equation}
Next scale  $\psi_0 =  \psi_{L^k}$ for $\psi$ on  $\bbT^{-k}_{N+M-k}$, and define
 $\cQ_{i}^{(k)}(\cA) =  \cQ_{i}(\cA_{L^{k-i}})\si_L^{k-i} $.  Then (\ref{form6})
is written 
\begin{equation} \label{form7}
\begin{split}
& \int \prod_{i=1}^{k} \cM_{\de \La_i,b_i}^{-1}   \exp \left(
  - b_i |\Psi_i- \cQ_{i}^{(k)}(\cA)\psi|^2_{\de \La_i}  \right) \exp \left( -
(\bpsi,(D_{e_k}(\cA )+m_k) \psi)\right)      f(\psi)  d \psi_{L^{-k}\La_0}  \\
\end{split}
\end{equation}
where  $f(\psi)   =   F(\psi_{L^k}) $.
Here  $\Psi_{0,\La^c}$ appears as   $\psi_{L^{-k} \La^c_0}   \equiv 
(\Psi_{0,\La_0^c})_{L^{-k}}$.

Next introduce   
\begin{equation}
\begin{split}
\bfpsi' = &( \Psi_{1,\de \La_1}, \cdots ,  \Psi_{k,\de \La_k})\\
\cQ_{k, \bla}(\cA) \psi  = &
 (( \cQ^{(k)}_1(\cA)  \psi)_{\de \La_1}, \cdots , (\cQ^{(k)}_k(\cA)  \psi)_{\de\La_k})\\
\end{split}
\end{equation}
and define 
\begin{equation}
( \overline \bfpsi' ,\bfpsi' )  =   \sum_{i=1}^k ( \bPsi_{i, \de \La_i},  \Psi_{i, \de \La_i} )
\end{equation}
We also define   $\bfb =  (b_1, \cdots , b_k)$ and 
      $  \cM_{\bla,\bfb} = \prod_{i=1}^{k}  \cM_{\de \La_i,b_i}^{-1} $.
Then  (\ref{form7}) is written  
\begin{equation}   \label{form8}
 \int  \cM^{-1}_{\bla,\bfb}  \exp \left(
  -  (\overline \bfpsi' - \cQ_{k, \bla} (-\cA) \bpsi, 
\bfb( \bfpsi' - \cQ_{k, \bla} (\cA)) \psi)
\right) 
\exp \left( -(\bpsi,(D_{e_k}(\cA )+m_k) \psi)\right) f(\psi) d \psi_{L^{-k}\La_0}  
\end{equation}

Again we diagonalize the quadratic form. Let     
\begin{equation}
D_k^\#(\cA) =   D_{e_k}( \cA )+m_k  + \cQ_{k,\bla}(-\cA)^T \bfb\cQ_{k,\bla}(\cA)
\end{equation}
This operator is invertible on $L^{-k}\La_0$ under  certain assumptions on  $\cA$ and  $\bla$ which we
explain in the next section.
Assuming it  is invertible  we define with  
   $\bfpsi =  (\Psi_{0,\La^c},  \bfpsi')$  
\begin{equation}
\begin{split}
S_{k,\bla}(\cA) =& [ D_k^\#(\cA) ]_{L^{-k}\La_0}^{-1}\\
\cH_{k,\bla}(\cA) \bfpsi=&  S_{k,\bla}(\cA)
\left(\cQ_{k,\bla}(-\cA)^T \bfb\bfpsi'   -   [ D_{e_k}(\cA)]_{\La, \La^c}
\psi_{\La^c} \right) ||_{\La =   L^{-k}\La_0}  
\\
\cH_{k,\bla}(\cA)\overline \bfpsi=&  S_{k,\bla}(\cA)^T
\left(\cQ_{k,\bla}(\cA)^T \bfb  \overline  \bfpsi'   -  ( [ D_{e_k}(\cA)]_{\La^c, \La})^T
\bpsi_{\La^c} \right) ||_{\La =   L^{-k}\La_0}  
\\
 (\overline \bfpsi, \bfD_{k, \bla}(\cA) \bfpsi) 
=&
(\overline \bfpsi' \left[  \bfb - \bfb\cQ_{k,\bla}(\cA)  S_{k,\bla}(\cA)\cQ_{k,\bla}(-\cA)^T \bfb \right]
\bfpsi')\\ 
+&
( \bpsi_{\La^c},  
 [D_{e_k}(\cA)]_{\La^c, \La}  S_{k,\bla}(\cA)\cQ_{k,\bla}(-\cA)^T\bfb\bfpsi' )\\
+&
( \cQ_{k,\bla}(\cA)^T  \bfb  \overline \bfpsi' , S_{k,\bla}(\cA)
\left[D_{e_k}(\cA)]_{\La, \La^c} \right]\psi_{\La^c}) \\
+&
( \bpsi_{\La^c} , \left[  ( D_{e_k}( \cA )+m_k ) -
 [D_{e_k}(\cA)]_{\La^c, \La}  S_{k,\bla}(\cA)[D_{e_k}(\cA)]_{\La, 
\La^c} \right] \psi_{\La^c} )|_{\La =   L^{-k}\La_0}\\
\end{split}
\end{equation}
Now make the change of variables    
 $\psi_{L^{-k} \La_0} \to \psi_{ L^{-k} \La_0} + \cH_{k, \bla}(\cA) \bfpsi $.
Then  (\ref{form8}) can be written   
\begin{equation}  \label{form9}
     Z_{k,\bla}(\cA) \ \exp \left(-  (\overline \bfpsi,   \bfD_{k, \bla}(\cA)  \bfpsi)\right) 
\int   f( \psi+\cH_{k, \bla}(\cA) \bfpsi )\  d\mu_{\st{S_{k,\bla}(\cA)}} 
(\psi_{L^{-k} \La_0})
\end{equation}
where  
\begin{equation} 
Z_{k,\bla}(\cA)  =\int 
 \cM_{\bla, \bfb}^{-1}\exp \left(-  ( \bpsi_{\La}, [S_{k, \bla}(\cA)]^{-1} \psi_{\La}) \right) 
 d \psi_{\La}  |_{\La = L^{-k} \La_0}
\end{equation}

Our final identity is then  (\ref{form 5.5}) = (\ref{form9}).  In case  $F=1$ it says 
\begin{equation}  \label{form 10}
\begin{split}
& \int   \prod_{j=0}^{k-1} d \Psi_{j, \La_j}   
 \cM_{L^{-1}\La_j',b}^{-1}  \exp \left(
  - \frac{b}{L} |\Psi_{j+1,L}- Q(\tilde \cQ_j\cA_{L^{k-j}} ) \Psi_{j}|^2_{\La_{j}}  \right)  \exp \left( -
(\bPsi_0,(D_{e_0}( \cA_{L^k} )+m_0) \Psi_0)\right) \\
&\ \ \ \ \ \ \ \ \ \ =    Z_{k,\bla}(\cA) \ \exp \left(- (\overline \bfpsi,  \bfD_{k, \bla}(\cA) 
\bfpsi)\right) 
\end{split}
\end{equation}

\section{Propagators}  \label{three}

\subsection{definitions}

In this chapter our goal is  to show that the propagators on  $\bbT^{-k}_{N+M-k}$
\begin{equation}
\begin{split}
G_{k, \bla}= &(-\De+ \mu^2_k  +   \cQ_{k,\bla}^T  \bfa   \cQ_{k,\bla} )_{L^{-k}\La_0 }^{-1}\\
S_{k,\bla}(\cA) =& (D_{e_k}( \cA )+m_k  + \cQ_{k,\bla}(-\cA)^T \bfb\cQ_{k,\bla}(\cA))_{L^{-k}\La_0 }^{-1}\\
\end{split}
\end{equation}
 exist and get estimates
on their kernels.    The basic tool is  a multiscale   random walk
expansion.   The expansion for Dirac operators was developed by  Balaban,  O'Carroll, and Schor \cite{BOS91}
  based on earlier work of  Balaban  \cite{Bal84}, \cite{Bal85}.   We give
expansions both for Dirac operators and Laplacians.    For Dirac operators  we follow  
\cite{BOS91} rather closely, but     nevertheless have to   go into considerable detail to 
get results in the exact  sharp form we want.   (In addition the numerous misprints in  \cite{BOS91} make it
difficult to quote results directly.)

We start with some definitions.
To construct the inverses 
we need to respect the structure of the averaging operators $ \cQ_{k,\bla} = \{ \cQ^{(k)}_i(\cdot)|_{\de
\La_i}\}$.  Now  $ ( \cQ_i \cA)_{\de \La_i}$
depends on  $\cA$ in the set  $L^{-i}B_i\de \La_i$ and hence the scaled version $ ( \cQ^{(k)}_i \cA)_{\de
\La_i}$ depends on  $\cA$ in the set  $\de \La_i^{(k)}$ given by 
\begin{equation}
\de \La_i^{(k)}  =   L^{-k} B_i \de \La_i  =
\left\{ \begin{array}{rcl}  
 L^{-k} (\La_{0}  - B_{1}\La_1) &     &  i=1 \\
 L^{-k} (B_{i-1}\La_{i-1}  - B_{i}\La_i) &     & 1< i <  k \\
 L^{-k} B_{k-1}\La_{k-1}   &     &  i =k \\
\end{array}
\right.
\end{equation}
Here for   $\La_i \subset  \bbT^{0}_{N+M-i}$    the set   $B_i\La_i  \subset   \bbT^{0}_{N+M}$ 
is its representative in   the original unit lattice, which is then scaled down to 
 $ L^{-k} B_{i}\La_i \subset  \bbT^{-k}_{N+M-k}$.
Since  we have assumed that  $\La_i$ is a union of  $LM_0$ blocks,   
   $B_i \La_i$ is a union of   $L^{i+1}M_0$ blocks,    $ L^{-k} B_i  \La_i $
is a union of $L^{-k+i+1}M_0$ blocks,  and  $\de \La_i^{(k)}$ is 
a union of  $L^{-(k-i)}M_0$ blocks.
The separation condition (\ref{separation}) 
insures that $\de \La_i^{(k)} $ is at least a few layers wide.

We have  the decomposition
of   $L^{-k}\La_0 \subset    \bbT^{-k}_{N+M-k}$ given by  the disjoint union
\begin{equation}
L^{-k}\La_0 =   \bigcup_{i=1}^k  \de \La_i^{(k)}  
\end{equation}
Let  $\bbD^0_i$ be the $L^{-(k-i)}M_0$ blocks in  $ \de \La_i^{(k)}  $ denoted $\square$, 
and let    $\bbD^0 = \cup_{i=1}^k  \bbD^0_i$.  Then
\begin{equation} \label{part1}
L^{-k}\La_0   =  \bigcup_{i=1}^k  \bigcup_{\square \in \bbD^0_i} \square=   \bigcup_{\square \in \bbD^0}
\square
\end{equation}
which is a partition of $L^{-k}\La_0 $ into blocks of various sizes.
Actually it is convenient to modify this by taking
 (for  $i \geq 2$ ) $r_0$ layers of $L^{-(k-i)}M_0$ blocks in $\de \La_i^{(k)}$ along the boundary 
of  $L^{-k} B_{i-1} \La_{i-1}$ and further subdividing them into  
 $L^{-(k-i+1)}M_0$ blocks.  We assume that  $r_0$ is some fraction of $r$
so that we do not exhaust $\de \La_i^{(k)}$.
Let  $\bbD_i$ be the new   $L^{-(k-i)}M_0$  blocks.  These are either in  $\de \La^{(k)}_i$
or   $\de \La^{(k)}_{i+1}$ and we have   
\begin{equation} \label{part2}
L^{-k}\La_0   =  \bigcup_{i=1}^k  \bigcup_{\square \in \bbD_i} \square=   \bigcup_{\square \in \bbD}
\square
\end{equation}

We will need partitions of unity concentrated on  the  sets  $\square  \in \bbD$.
First take a smooth function  $g$  on $\bbR^3$ so   $g$ has support in  $\{x: |x| \leq 2/3\}$ and $g=1$
on   
$\{x: |x| \leq 1/3\}$      and   $\sum_{n \in \bbZ^3}  g(x-n)^2  =1$.  
Then if   $ \square $ is a   $\bbD_i$ block  in the interior of   $\bigcup_{\square \in \bbD_i} \square$ 
centered on $y
\in
\bbT^{-(k-i)+m_0}_{N+M-k}$, we define 
\begin{equation}
h_{\square} (x)   =   g\left( (x-y) \frac{ L^{k-i}}{M_0}\right)
\end{equation} 
Then for  $x$ well inside $\bigcup_{\square \in \bbD_i} \square$
\begin{equation}   \label{sum}
\sum_{\square  \in \bbD}  h_{\square} (x)   ^2 =1 
\end{equation}
For  $ \square \in  \bbD_i$ blocks touching  $\bbD_{i-1}$ blocks  the scalings do not match.
We require in this case that 
the definition of  $h_{\square} (x) $ be modified
on any boundary face by  taking the scaling  $ L^{k-(i-1)}/M_0$ instead of  $ L^{k-i}/M_0$. 
(If $\square \subset \de \La_1^{(k)}$ touches $\La_0^c$  take  $ L^k/3M_0$.)     Then 
(\ref{sum} ) holds for all  $x \in  L^{-k}\La_0$. 

Note that   $h_{\square}h_{\square'} =0$ unless  $\square, \square'$ touch.

We also define some enlargements of each  $L^{-(k-i)}M_0$ cube  $\square \in \bbD_i$.
We set  
\begin{equation}
\begin{split}
\tilde \square   =&   3M_0L^{-(k-i)} \textrm{ cube centered on } \square \\
\square^{(n)} = &  (1+2n)r_1M_0 L^{-(k-i)}  \textrm{ cube centered on } \square \\
\end{split}
\end{equation}
for some $r_1 <r_0$.
We have    $supp \ h_{\square}  \subset  \tilde \square$.

Finally we introduce some   modified distances  on $\bbT^{-k}_{N+M-k}$
For long distances we use
\begin{equation}
d_{\bla}(x,y)  =  \inf_{\Ga: x \to y} \sum_{i=1}^k  L^{k-i} |\Ga \cap \de \La_i^{(k)}|
\end{equation} 
This   is a genuine metric which  weighs earlier regions more heavily.
For short distances we use   
\begin{equation}
   d'(x,y)  = \left\{
\begin{array}{rl}
 d(x,y)  &  x\neq y \\
L^{-k}   & x=y\\  
\end{array}  \right.
\end{equation}
This  is not a
real metric but it does satisfy the triangle inequality, and it does scale like $d(x,y)$.   It is relevant because of
its appearance in estimates like (\ref{uno}),(\ref{duo}).

We note the following estimates on integrals in  $\bbT^{-k}_{N+M-k}$.  As usual 
$\int dy[\cdots] =  \sum_y L^{-3k} [\cdots]$. 
The estimates refer to   $L^{-(k-i)}$ blocks  $\De$, smaller than the  
$L^{-(k-i)}M_0$ blocks  $\square$.

\begin{lem}
{ \ \ \  }
 Let   $\De$ be an  $L^{-(k-i)}$ block with  $1 \leq i \leq k$ and let     $0   \leq 
\al < 3$ 
\begin{eqnarray}  \label{estimate}
 \int_{ \De}  d'(x,y)^{-\al} dy& \leq & \cO(L^{-(3-\al)(k-i)} )  \label{e1}\\
\int_{ \De }d'(x,y)^{-\al} d'(y,z)^{-\al} dy & \leq & \cO(L^{-(3-\al)(k-i)} )d'(x,z)^{-\al}
\label{e2} \\
\int_{\De }d'(x,y)^{-1} d'(y,z)^{-2} dy& \leq & \cO(L^{-(k-i)} )d'(x,z)^{-1} 
\label{e3}
\end{eqnarray}
\end{lem}
\bigskip

\pr  For the first estimate note that if $x$ is well outside  $\De$ then the integrand is 
bounded and we get  $\cO(L^{-3(k-i)})$ which suffices.  If $x$ is in or near  $\De$ enlarge 
$\De$  so it is centered on  $x$ and still has sides  $\cO(L^{-(k-i)})$. The point with
$y=x$ contributes  $L^{-(3-\al)k}$ which suffices.  Points with  $y \neq x$ can be  dominated
by the $\bbR^3$ integral  $\int_{|y| \leq  \cO(L^{-(k-i)})} |y|^{-\al} dy  =  \cO(L^{-(3-\al)(k-i)} ) $.

For the second integral we split into two cases.  The first is $d'(x,y) \geq d'(x,z)/2$.  Use
this in the first factor and then  (\ref{e1}) gives the result.  
The second case is  $d'(x,y) < d'(x,z)/2$.  In this case we have by the 
triangle inequality that     $d'(y,z) \geq d'(x,z)/2$.  Use this in the second factor and again
use (\ref{e1}) to obtain the result.  

For the last inequality regard the integrand as the product of   $d'(x,y')^{-1} d'(y',z)^{-1} $
and $d'(y',z)^{-1}$ and use the Schwarz inequality.

\subsection{fermions}

The random walk expansion for   $S_{k, \bla} (\cA,x,y )$ has the 
form 
\begin{equation} \label{walk}
S_{k, \bla} (\cA,x,y )=   \sum_{\om: x \to y}  S_{k, \bla, \om }(\cA,x,y)
\end{equation}  
Here we are summing over paths $\om$  each of which is  a  sequence of adjacent cubes  (blocks) $\square_0,
\dots
\square_n$ from   $\bbD$.
Equivalently a path $\om$ is a sequence of links   $( (\square_0,\square_1) , (\square_1,\square_2),\dots , 
(\square_{n-1},\square_n)$  and the adjacency condition is that that $\square_j, \square_{j+1}$
should touch, possibly only on corners,  and including the possibility   $\square_j = \square_{j+1}$.
The notation     $\om: x \to y$ means  $x \in \tilde  \square_0, y \in  \tilde \square_n$. 
We let    $|\om| =n$ be the number of links.    If  $| \om| = 0$ then there is just
the single square  $\square_0$.

\begin{thm}  \label{f}
Let    $\cA$ satisfy    
\begin{equation}
|\pa \cA | 
\leq C L^{3(k-i)/2}p(e_i) \textrm{  on  }\de\La_i^{(k)}
\end{equation}
 for some constant $C$ and   $p(e_i)= \log(e_i^{-1})^p$. Let
  $M_0$ be sufficiently large and  let  $e_k$ be   sufficiently small.
Then  
$S_{k, \bla} (\cA,x,y )$ exists and has the random walk expansion   (\ref{walk}).
We have the bound for each path
\begin{equation}  \label{a}
\begin{split}
| S_{k,\bla,\omega}(\cA, x,y)  |
\leq & \cO(1) (\cO(1)M_0^{-1} )^{|\om|}d'(x,y)^{-2} \exp( - \cO(1) d_{\bla}(x,y) ) \\
\end{split}
\end{equation}
and the bound for the full propagator
\begin{equation}   \label{b}
\begin{split}
|S_{k, \bla } (\cA,x,y )
\leq  & \cO(1) d'(x,y)^{-2}  \exp( - \cO(1) d_{\bla}(x,y) ) \\
\end{split}
\end{equation}
In addition  $S_{k, \bla, \om   }(\cA)$  and  $S_{k, \bla } (\cA,x,y )$  are  gauge covariant,
 and   $S_{k, \bla, \om   }(\cA)$  depends
on 
$\cA$ only in 
$\bigcup_{\square \in \om} \square^{(5)}$. 
\end{thm}
\bigskip

 \res
\begin{enumerate}
\item  In  $ \exp( - \cO(1) d_{\bla}(x,y) ) $ the $\cO(1)$ is independent of  $M_0$.
\item   There is no condition on $\cA$ itself, but only on  $\pa \cA$.  
  This important feature follows from the 
gauge covariance as we will see. (And  actually a condition
on the field strength $F= d \cA$ would suffice.)
\item  It is possible that   $\bla$ is a sequence of  the full tori, i.e.  $\La_i = \bbT^0_{N+M-i}$.  
In this case   $d_{\bla}(x,y) =d(x,y)$ and   we have  the  result  (\ref{duo}) for  $S_k(\cA,x,y)$.
\end{enumerate}

\bigskip

To prove the theorem  we need a result for a single block.

\begin{lem}   \label{lem}
 Under the same hypotheses for each $\bbD_i$ block     $\square$    there is 
an operator     $S^*_{\square}(\cA )=S^*_{k,\square}(\cA) $ such that  for    $x \in  \tilde  \square $ 
\begin{equation}  \label{fence}
 \left( \left(D_{e_k}( \cA )+m_k  + \cQ_{k,\bla}(-\cA)^T  \bfb \
\cQ_{k,\bla}(\cA)\right) S^*_{\square}(\cA)f\right)(x)  = f(x)  
\end{equation}
and for all $x,y \in \tilde  \square$
\begin{equation}  \label{star}
 |S^*_{\square}(\cA,x,y)| \leq     \cO(1)  d'(x,y)^{-2}\exp( - \cO(1) d_{\bla}(x,y) )  
\end{equation}
In addition  $ S^*_{\square}(\cA)$ is gauge covariant and   depends on  $\cA$  only in $\square^{(5) }$
\end{lem}
\bigskip

Assuming the lemma we have
\bigskip

\noindent
\textbf{Proof of theorem \ref{f}.}  

\noindent
\textbf{Part I.  } We  define a parametrix on $L^{-k}\La_0$
\begin{equation}
S^*(\cA)   =   \sum_{\square \in \bbD} h_{\square}   S^*_{\square}(\cA)  h_{\square}
\end{equation}
Then thanks to  (\ref{sum}) and  (\ref{fence})
\begin{equation}  
\left( D_{e_k}( \cA )+m_k  + \cQ_{k,\bla}(-\cA)^T \bfb \ 
\cQ_{k,\bla}(\cA) \right) S^*(\cA)  =  I - \sum_{\square}
R_{\square}(\cA)    S^*_{\square}(\cA)    h_{\square} \equiv I -R 
\end{equation}
where  
\begin{equation}
 R_{\square}(\cA) = - \left[ \left( D_{e_k}( \cA )+m_k  + \cQ_{k,\bla}(-\cA)^T \bfb \ 
\cQ_{k,\bla}(\cA)\right), h_{\square}\right]
\end{equation}
The inverse is now 
\begin{equation}   \label{Neumann}
S_{k,\bla} (\cA)   = S^*(\cA)   (I - R)^{-1} 
=   S^*(\cA)   \sum_{n=0}^{\infty} R^n  
\end{equation}
provided the series  converges.
This can also be written 
\begin{equation}  \label{frandom}
\begin{split}
S_{k, \bla}(\cA) =& \sum_{n=0}^{\infty}  \sum_{ \square_0, \square_1,...,\square_n}
\left( h_{\square_0} S^*_{\square_{0}}(\cA)    h_{\square_{0}}\right)
\left(R_{\square_1}(\cA)    S^*_{\square_{1}}(\cA)    h_{\square_{1}}\right)
\cdots 
\left(R_{\square_n}(\cA)    S^*_{\square_{n}}(\cA)    h_{\square_{n}}\right)\\
\equiv & \sum_{\om}   S_{k,\bla,\omega}(\cA)  \\
\end{split}
\end{equation}
In the last step we identify the 
 random walk expansion by noting  that the term vanishes unless all pairs  $\square_j,\square_{j+1}$
are adjacent.  The kernel   $S_{k,\bla,\omega}(\cA,x,y)$ vanishes unless
$ x \in supp\ h_{\square_0} \subset \tilde  \square_0$  and   $ y \in supp\ h_{\square_n} \subset
\tilde  \square_n$.
The gauge covariance of   $ S_{k,\bla,\omega}(\cA)  $ follows from that of   
$  S^*_{\square}(\cA)   $ and   $R_{\square}(\cA)$, as does the $\cA$ dependence.
 \bigskip

\noindent
\textbf{Part II.  } 
To estimate this expansion we need the following bound.  For  $\square \in \bbD_i$
\begin{equation}  \label{fkey}
|(R_{\square}  (\cA)  S^*_{\square} (\cA)) (x,y)| 
\leq  \cO\left(\frac{L^{k-i}}{M_0}\right)    d'(x,y)^{-2}\exp( - \cO(1) d_{\bla}(x,y) ) 
\end{equation}
To prove it we write  $R_{\square} = R^D_{\square} +R^Q_{\square}$  where    $R^D_{\square}
=   - [  D_{e_k}, h_{\square}]$  and    $R^Q_{\square}= -[  \cQ_{k,\bla}^T \bfb
\cQ_{k,\bla}, h_{\square}] $  .
We have explicitly
\begin{equation}
   (R^D_{\square}  (\cA)   S^*_{\square} (\cA)) (x,y) 
= \sum_{x'}
\ga_{x, x'} L^k  e^{ie_kL^{-k} \cA(x,  x')}  (h_{\square}(x) - h_{\square}( x')) S_{\square}^*( x',y)
\end{equation}
where the sum is over nearest neighbors  $x'$ of $x$.
The result now follows by 
\begin{equation}
L^k |h_{\square}(x) - h_{\square}( x')| \leq \sup_x|\pa h_{\square}(x)|  \leq 
\cO\left(\frac{L^{k-i}}{M_0}\right)  
\end{equation}
 and (\ref{star}) and 
  $d'(x',y) \geq  d'(x, y)/2$    and   $d_{\bla}(x',y) \geq d_{\bla}(x,y)-1$.

For the other term  we have from   (\ref{explicit}) on  $\bbT^{-i}_{N+M-i}$ 
\begin{equation}
(\cQ_i(-\cA)^T b_i\cQ_i(\cA)))(x,x')
=    \exp\left(ie_i \cA(\tilde \Ga_{x,[x]} \cup \tilde \Ga_{[x],x'})\right)
\chi(|x'-[x]| \leq 1/2  )  
\end{equation} 
where   $[x]$ is the unit lattice point at the center of 
the block containing  x.  Then for the scaled version on    $\bbT^{-k}_{N+M-k}$   
\begin{equation}
\begin{split}
(\cQ^{(k)}_i(-\cA)^T b_i\cQ^{(k)}_i(\cA))(x,x')
=& L^{2(k-i)} (\cQ_i(-\cA_{L^{k-i}})^T b_i\cQ_i(\cA_{L^{k-i}}))(L^{k-i}x, L^{k-i}x')  \\
=& L^{2(k-i)} b_i \exp\left(ie_i\cA_{L^{k-i}}(\dots)\right)
\chi(|x'-[x]| \leq L^{-(k-i)}/2  )   \\
\end{split}
\end{equation}
where now   $[x]$ is the  $L^{-(k-i)}$
lattice point at the center of  the  $L^{-(k-i)}$ block containing $x$.
Now  $\square  \subset \de \La_i^{(k)} \cup \de  \La_{i+1}^{(k)}$
and the same is true for   $supp \ h_{\square}  \subset \tilde \square$.
Hence the only contribution to   $R_{\square}^Q$ comes from
\begin{equation}
\left( \cQ_{k,\bla}(-\cA)^T \bfb \ \cQ_{k,\bla}(\cA) \right)(x,x')
=  \left\{   \begin{array}{rcl}
(\cQ^{(k)}_i(-\cA)^T b_i\cQ^{(k)}_i(\cA))(x,x')&  &  x,x' \in  \de \La_i^{(k)} \\
(\cQ^{(k)}_{i+1}(-\cA)^T b_{i+1}\cQ^{(k)}_{i+1}(\cA))(x,x') &  &  x,x' \in  \de \La_{i+1}^{(k)} \\
\end{array}   \right.
\end{equation}
We concentrate on the first case, the other is similar.
Then  
\begin{equation}
 (R^Q_{\square}   (\cA)  S^*_{\square} (\cA)) (x,y) 
=  b_i \int_{|x'-[x]| \leq L^{-(k-i)}/2} 
L^{2(k-i)}    \exp\left(ie_i\cA_{L^{k-i}}(\dots)\right)(h_{\square}(x)- h_{\square}( x')) S^*_{\square}
(x',y) 
\end{equation}
Now  since  $|x-x'|  \leq   L^{-(k-i)}$
\begin{equation}
 |h_{\square}(x) - h_{\square}( x')| \leq L^{-(k-i)} \sup_x|\pa h_{\square}(x)|  \leq  \cO(M_0^{-1})  
\end{equation}
we obtain
\begin{equation}
\begin{split}
| (R^Q_{\square} (\cA)    S^*_{\square} (\cA)) (x,y)| 
\leq  & \cO\left(\frac{L^{2(k-i)}}{M_0}\right)    \int_{|x'-[x]| \leq L^{-(k-i)}/2} 
   d'(x',y)^{-2} \exp( - \cO(1) d_{\bla}(x,y) )  \\
\leq  &  \cO\left(\frac{L^{k-i}}{M_0}\right)  \exp( - \cO(1)
d_{\bla}(x,y) )  \\ 
\end{split}
\end{equation}
In the second step we use (\ref{e1}).
Since   $d'(x,y) \leq  d_{\bla}(x,y)+1$ the result follows.
\bigskip

\noindent
\textbf{Part III.  } 
Now we estimate the expansion. Besides the partition  (\ref{part1}) we can
also partition $L^{-k}\La_0$ into smaller blocks  $\De$ of size   $L^{-(k-i)}$ in   $\de \La_i^{(k)}$.
Then   for   $\om  = (\square_0,\dots,\square_n)$
\begin{equation}  
\begin{split}
S_{k, \bla, \om }(\cA,x,y) =&  \sum_{\De_1, \dots,\De_n}
  \int_{\De_1}dx_1 \dots  \int_{\De_n}dx_n\\
 &
( h_{\square_0} S^*_{\square_{0}} (\cA)    h_{\square_{0}})(x,x_1)
  (R_{\square_1}    S^*_{\square_{1}} (\cA)    h_{\square_{1}})(x_1,x_2)
\cdots 
 (R_{\square_n}    S^*_{\square_n}   (\cA)  h_{\square_n})(x_n,y)\\
\end{split}
\end{equation} 
We can restrict the sum to $\De_j$  intersecting both  $\tilde \square_{j-1}$ and    
$\tilde  \square_{j}$.  Now use the estimates (\ref{star}),  (\ref{fkey}) to obtain 
\begin{equation}  
\begin{split}
&|S_{k, \bla\om}(\cA,x,y) |\leq    (\cO(1)M_0)^{-n})\sum_{\De_1, \dots,\De_n}
    \int_{\De_1}dx_1 \dots  \int_{\De_n}dx_n\\
 &
 \left( d'(x,x_1)^{-2}e^{ - \cO(1) d_{\bla}(x,x_1) }  \right) 
 \left( L^{k-i_1}  d'(x_1,x_2)^{-2}e^{ - \cO(1) d_{\bla}(x_1,x_2) } \right)
\cdots 
 \left(L^{k-i_n}  d'(x_n,y)^{-2}e^{ - \cO(1) d_{\bla}(x_n,y) } \right)\\
\end{split}
\end{equation} 
Here  $i_j$  is chosen by   $\square_j \in  \bbD_{i_j}$. 
Now use  
$d_{\bla}(x_i,x_{i+1})  \geq   d_{\bla}(\De_i,\De_{i+1}) -2$ where the distance is 
from the center of the cubes. 
Then   repeatedly use  the estimate
\begin{equation}    \label{est10}
  L^{k-i_j} \int_{\De_j}  d'(x_{j-1},x_j)^{-2}
  d'(x_j,x_{j+1})^{-2}  dx_j \leq    \cO( d'(x_{j-1},x_{j+1})^{-2})
\end{equation}
which follow from  (\ref{e2}).  Here we use that  $\tilde \square_j$ is contained in
$\de \La_{i_j} \cup   \de \La_{i_j+1}$, hence so is $\De_j$ and hence it 
is either a  $L^{-(k-i_j)}$ or a   $L^{-(k-i_j-1)}$ block.  
We also use the estimate  
\begin{equation}   \label{est11}
\sum_{\De_1,\dots,\De_n}
 e^{-  \cO(1) d_{\bla}(x,\De_1)}e^{-  \cO(1) d_{\bla}(\De_1,\De_2)}\dots e^{-  \cO(1) d_{\bla}(\De_n,y)}
\leq   (\cO(1))^n    \exp\left(-  \cO(1) d_{\bla}(x,y)\right) 
\end{equation}
For this see \cite{Bal84}, lemma 2.1.
 These estimates yield the bound on
$S_{k, \bla,\om}(\cA,x,y)$.   For the bound on  $S_{k, \bla}(\cA,x,y)$  we sum over paths. 
The factor  $(\cO(1)M_0)^{-n}  $ is sufficient to control the sum if 
$M_0$   is   sufficiently  large.
\bigskip

\noindent
\textbf{proof of lemma  \ref{lem}}
\bigskip

\noindent
\noindent

\noindent
\textbf{part I.}
We need to invert $\left(D_{e_k}( \cA )+m_k  + \cQ_{k,\bla}(-\cA)^T \bfb
\cQ_{k,\bla}(\cA)\right)$ on a single block. 
Inverting  with straight  Dirichlet boundary conditions 
makes it awkward to get estimates, so we use  a kind of soft Dirichlet conditions, 
following  \cite{BOS91} and  the construction  of the theorem.

Given  $\bla$ we take a fixed  $\bbD_i$ cube  $\square \subset 
T^{-k}_{N+M-k}$.   
Then  define a  decreasing   sequence of cubes
\begin{equation}
\bom(\square)  =  ( \Om_0(\square), \dots,  \Om_{k-1}(\square) )
\end{equation}
with $\Om_j(\square)  \subset   \bbT^0_{N+M-j}$.
In reverse order they are specified by 
\begin{equation}
\begin{split}
\Om_{j}  =& \emptyset    \ \ \ \ \ \ \    j>i  \\
  \Om_{i}(\square) =&  U_iL^k \square ^{(2)} \cap  \La_{i}\\
  \Om_{i-1}(\square) =&  U_{i-1} L^k \square^{(3)}    \\
 d(( L^{-1}\Om'_{j-1}(\square))^c,  \Om_{j}(\square) ) =& r_1M_0  \ \ \ \ \ \ j=i-1 \dots  1\\
\end{split}
\end{equation}
where  the cube  $L^{-1}\Om'_{j-1}(\square)$  is required to be centered on  $\Om_{j}(\square)$. 
(If $i=k$ then  $\Om_i(\square) = \emptyset$.)

Now with      $\de \Om_j(\square)  =  L^{-k} (B_{j-1}\Om_{j-1}(\square)  -  B_j \Om_j(\square) )$  we have 
for   $j=1, \dots,k$
\begin{equation}
\begin{split}
\de  \Om^{(k)}_{j}(\square)  =& \emptyset    \ \ \ \ \ \ \    j>i+1  \\
 \de  \Om^{(k)}_{i+1}(\square) =&    \square ^{(2)} \cap L^{-k}B_i \La_{i}\\
 \de  \Om^{(k)}_{i}(\square) =&  \square^{(3)}  -(  \square ^{(2)}\cap L^{-k}B_i \La_{i})\\
d(   L^{-k} (B_{j-1}\Om_{j-1}(\square))^c,  L^{-k} B_j \Om_j(\square) )   \leq  &  L^{-(k-j)}
r_1M_0 
\\  
\end{split}
\end{equation}
Here we use that   $L^k \square ^{(3)} $ is a $7r_1M_0L^i$ cube and so has the form  
$B_i \tilde \square$ for some  $7r_1M_0$ cube  $\tilde \square$ and hence   $B_iU_i =I$ on this set.
  We have also used 
$d(B_1X,B_1Y) \leq L d(X,Y)$ and $(B_1X)^c = B_1 X^c$.
We note that
\begin{equation}
d(L^{-k} \Om_0(\square)^c,  \square^{(3)})  \leq   \sum_{j=1}^i    L^{-(k-j)}
r_1M_0 \leq  2 r_1M_0 L^{-(k-i)}
\end{equation}
 which implies that    $L^{-k} \Om_0(\square)  \subset  \square^{(5)}$.

We  now    define   
\begin{equation}  \label{sstr}
\begin{split}
S^*_{\square}(\cA) =  &  S_{\bom(\square)}(\cA)
= \left[ D_{e_k}( \cA )+m_k  + \cQ_{\bom(\square)}(-\cA)^T
\bfb \cQ_{\bom(\square)}(\cA) \right]_{L^{-k}\Om_0(\square)}^{-1}\\
\end{split}
\end{equation}
if it exists.

Before considering existence we prove that    
(\ref{sstr}) satisfies (\ref{fence}). It suffices to show that  
for    $x \in \tilde  \square$ 
\begin{equation}   \label{agree}
\left(\cQ_{\bom(\square)}(-\cA)^T
\bfb \cQ_{\bom(\square)}(\cA)f\right)(x)
=  \left( \cQ_{k,\bla}(-\cA)^T  \bfb \
\cQ_{k,\bla}(\cA) f\right)(x)
\end{equation}
Recall  that  $\tilde \square
\subset \de \La_i^{(k)}  \cup  \de \La_{i+1}^{(k)}$.
If   $x \in   \tilde \square\cap  \de \La_{i+1}^{(k)}$ then   
$x \in   \tilde \square\cap  L^{-k}B_i \La_i  \subset \de \Om_{i+1}(\square)$ 
and the left side  of  (\ref{agree}) is      $\left( \cQ^{(k)}_{i+1}(-\cA)^T  b_{i+1} \
\cQ^{(k)}_{i+1}(\cA) f\right)(x)$  which agrees with the right side of  (\ref{agree}).
If   $x \in   \tilde \square\cap  \de \La_{i}^{(k)}$ then   
$x \in   \tilde \square -  L^{-k}B_i \La_i  \subset \de \Om_{i}(\square)$ 
and the left side  of  (\ref{agree}) is      $\left ( \cQ^{(k)}_{i}(-\cA)^T  b_{i} \
\cQ^{(k)}_i(\cA) f\right)(x)$  which agrees with the right side of  (\ref{agree}).
Thus (\ref{agree}) is true.
\bigskip

\noindent
\textbf{part II.} We are going to treat  $S^*_{k,\square}(0)$  first and we 
start with some definitions at  $\cA =0$.   Let  $S_j = S_j(0)$  and 
 $\cQ_j= \cQ_j(0)$ and 
 $S_j^{(k)}=  \si^{-1}_{L^{k-j}} S_j  \  (\si^{-1}_{L^{k-j}})^T$
which is the representation of  $S_j$ on  $\bbT^{-k}_{N+M-k}$.
We have  
\begin{equation}
\begin{split}
    S_j^{(k)}    \equiv &\ \si_{L^{k-j}}^{-1}  (\ D(0)+m_j + \cQ_{j}^T b_j\cQ_{j}\ )^{-1}
 \  (\si^{-1}_{L^{k-j}})^T \\
=&   (\ D(0) +m_k + \cQ_j^{(k)T}  b_j\cQ_{j}^{(k)}\ )^{-1}\\
\end{split}
\end{equation}
and   the bound   from (\ref{duo})
\begin{equation}   
| (S_j^{(k)})(x,y)| =
|L^{2(k-j) }S_j(L^{k-j}x,L^{k-j}y)|
\leq 
\cO(1) d'(x,y)^{-2}  \exp( - \cO(1) L^{k-j}d(x,y) )
\end{equation}

We also consider the mixed operator for suitable  $Y \subset \bbT^{-k}_{N+M-k}$
\begin{equation}
\begin{split}
& [S_{j}^{(k)} ]_Y = \left(\ D(0) +m_k +[  \cQ_j^{(k)T}b_j\cQ_{j}^{(k)}\ ]_ {Y^c}\
+ [  \cQ_{j+1}^{(k)T} b_{j+1} \cQ_{j+1}^{(k)}\ ]_Y   \right)^{-1}\\
\end{split}
\end{equation}
This has the alternate representation from  \cite{BOS91} .
\begin{equation}   \label{mixed}
\begin{split}
[S_j^{(k)}]_Y 
=& \si_{L^{k-j}}^{-1}  \left( S_{j}  +b_j^2 S_{j} \cQ_{j}^T 
[\ D_j(0)  + \frac{b}{L} Q^TQ  ]_{U_jL^kY}^{-1} \cQ_{j} S_{j} \right)
(\si_{L^{k-j}}^{-1} )^T\\
&  S_{j}^{(k)}  + b_j^2S_{j}^{(k)} \cQ_{j}^{(k)T} 
[\ D_j(0)  + \frac{b}{L} Q^TQ  ]_{U_jL^kY}^{-1} \cQ_{j}^{(k)} S_{j}^{(k)} \\
\end{split}
\end{equation}
To estimate this we  take the bound   from  \cite{BOS91}  for  $z,w \in \bbT^0_{N+M-j}$
\begin{equation}
|\left[\ D_j(0)  + \frac{b}{L} Q^TQ  \right]_X^{-1}(z,w)| 
\leq  \exp(- \cO(1) d(z,w) )
\end{equation}
Then  the expression in parentheses in 
(\ref{mixed}) has a kernel which is $\cO(1) d'(x,y)^{-2}  \exp( - \cO(1) d(x,y) )$.  
(Even without the short distance singularity for the second term.)  Thus 
after scaling we have again
\begin{equation}   \label{mixedbound}
| [S_j^{(k)}]_Y (x,y)| 
\leq 
\cO(1) d'(x,y)^{-2} \exp( - \cO(1) L^{k-j}d(x,y) )
\end{equation}
\bigskip

\noindent
\textbf{part III.}
Now we show  $S^*_{\square}(0)$ exists and get an estimate on the kernel.  
  First  define a parametrix  on  $L^{-k}\Om_0(\square)$
\begin{equation}
S^\#(\square)    =   \sum_{\square' \subset  \bbD(\square)  } h_{\square'}   S^\#_{\square'} (\square)
h_{\square'}
\end{equation}
Here   $\bbD(\square)= \cup_j \bbD_j(\square)  $  is a     partition of   of   $L^{-k}\Om_0(\square) =  
\cup_j\de\Om_j^{(k)}(\square) $ 
 defined by dividing   $\de\Om_j^{(k)}(\square)$
 into cubes  of size 
$L^{-(k-j)}M_0$ and then further subdividing cubes touching   $\de\Om_{j-1}^{(k)}(\square)$. 
Then  $\bbD_j(\square)$ is the set of  new   $L^{-(k-j)}M_0$  cubes.  These are contained in   $\de \Om_j^{(k)}
\cup  \de \Om_{j+1}^{(k)}$.
   We define    
  $S^\#_{\square'}(\square)$  as follows.
If   $\square'  \in  \bbD_j(\square)$ and $(\square')^{(2)}$ does not intersect
$ L^{-k}B_{j}\Om_{j}(\square)$ then we define  $ S^\#_{\square'}(\square)=  S_j^{(k)}  $.  More
generally   we define 
\begin{equation}
   S^\#_{\square'}(\square)= [ S_j^{(k)} ]_Y \ \ \ \ \ \ \              Y =    (\square')^{(2)}  \cap 
 L^{-k}B_{j}\Om_{j}(\square)
\end{equation}
This use of whole lattice inverses  would not be suitable for  $\cA \neq 0$ since it would not be local in $\cA$.

 Then   $S^\#_{\square'}(\square)$ provides a local inverse in the sense that for  $x \in  \tilde \square'$
\begin{equation}
\left(  \left (D_{e_k}(0 )+m_k  + \cQ_{\bom(\square)}(0)^T
\bfb \cQ_{\bom(\square)}(0) \right)   S^\#_{\square'}(\square) f  \right)(x) =f(x)
\end{equation}
Check this as in  (\ref{agree}).
If  $ \square' \in \bbD_j$ and   $x,y \in \tilde \square'$   
we have from   (\ref{mixedbound})  
\begin{equation}  
|S^\#_{\square'}(\square)(x,y)| \leq   \cO(1) d'(x,y)^{-2}  \exp( - \cO(1) d_{\bom(\square) }(x,y) )
\end{equation}

Now we compute  
\begin{equation}  
( D_{e_k}( 0)+m_k  + \cQ_{\bom(\square)}^T \bfb
\cQ_{\bom(\square)}) S^\#(\square)  =  I - \sum_{\square'}
R^\#_{\square'} (\square)   S^\#_{\square'}(\square)    h_{\square'} \equiv I -R ^\#(\square)
\end{equation}
where  
\begin{equation}
 R^\#_{\square'}(\square) = - \left[ (D_{e_k}( 0)+m_k  + \cQ_{\bom(\square)}^T \bfb\ 
\cQ_{\bom(\square)}), h_{\square'}\right]
\end{equation}
The inverse  on   $L^{-k}\Om_0(\square)$ is then 
\begin{equation}  
S^*_{\square}(0)   = S^\#(\square)   (I - R^\#(\square))^{-1} 
=   S^\#(\square)   \sum_{n=0}^{\infty} (R^\#(\square))^n  
\end{equation}
if it converges.
This can also be written 
\begin{equation}
\begin{split}
S^*_{\square}(0) =& \sum_{n=0}^{\infty}  \sum_{ \square_0, \square_1,...,\square_n}
( h_{\square_0} S^\#_{\square_{0}}(\square)    h_{\square_{0}})
(R^\#_{\square_1}(\square)    S^\#_{\square_{1}}(\square)    h_{\square_{1}})
\cdots 
(R^\#_{\square_n}(\square)    S^\#_{\square_{n}} (\square)   h_{\square_{n}})\\
\end{split}
\end{equation}
Convergence is demonstrated just as before and gives 
\begin{equation}  \label{same}
|S^*_{\square}(0,x,y)| \leq   \cO(  d'(x,y)^{-2}  )    \exp\left( -  \cO(1) d_{\bom(\square)}(x,y)\right) 
\end{equation}
Now consider  $x,y \in L^{-k}\Om_0(\square) \subset \square^{(5)}$.  Assuming $5r_1 < r_0$ we have 
$\square^{(5)} \subset  \de \La_i^{(k)} \cup  \de \La_{i+1}^{(k)}$.  Then   
\begin{equation}  \label{so}
 d_{\bom(\square)}(x,y)  \geq   \cO(1)L^{(k-i)} d(x,y)   \geq   \cO(1) d_{\bla}(x,y)
\end{equation}
and we have the result  (\ref{star})  we need.
\bigskip

\noindent
\textbf{part IV.}  We continue to consider  $\square  \in  \bbD_i$
and now   study   $S_{\square}^*(\cA)$ for    $\cA \neq 0 $.
At first suppose that instead of the  bound on  $\pa \cA$  we have for some $C$
 \begin{equation}
|\cA|  \leq  C L^{(k-i)/2}p(e_i) \textrm{  on   }\de \La_i^{(k)}
\end{equation}
The same bound holds on  $\de \La_i^{(k)}  \cup  \de \La_{i+1}^{(k)}$ since  $p(e_{i+1})<p(e_i)$.
Hence the bound holds on 
$\tilde  \square^{(5)}$ and hence on   $L^{-k}\Om_0(\square)$, 
the region we are working in.

Let  
\begin{equation}
\begin{split}   v_{\square}(\cA,\cA') =&
\ [D_{e_k}(\cA) +m_k +  \cQ_{\bom(\square)}(-\cA)^T
\bfb \cQ_{\bom(\square)}(\cA)]\\
- &[D_{e_k}(\cA)' +m_k +   \cQ_{\bom(\square)}(-\cA')^T
\bfb \cQ_{\bom(\square)}(\cA')]\\
\end{split}
\end{equation}
Then    $S^*_{ \square}(\cA)$ exists if the series   
\begin{equation}  \label{monday}
S^*_{ \square}(\cA) =S^*_{ \square}(0)
\left( \sum_{n =0}^{\infty} \left(  v_{\square}(\cA,0)S^*_{ \square}(0) \right)^n
\right)
\end{equation}
converges.

To show convergence we need to estimate the kernel $(v_{\square}(\cA,0)S^*_{ \square}(0))(x,y)$.
Again there are two parts  coming from $v_{\square}(\cA,0) =  v_{\square}^D(\cA,0) +  v_{\square}^Q(\cA,0)$.
For the first term
\begin{equation}
 (v^D_{\square} (\cA,0)   S^*_{\square}(0)) (x,y) 
=    \sum_{ x'}
\ga_{x, x'} L^k ( e^{ie_kL^{-k} \cA(x, x')} -1)  ( S^*_{\square}(0))( x',y) 
\end{equation} 
which is bounded by   
\begin{equation}
\begin{split}
| (v^D_{\square} (\cA,0)   S^*_{\square}(0)) (x,y) |
\leq  &  \cO(1) e_k    
 \sup| \cA |\ d'(x,y)^{-2} \exp( - \cO(1) d_{\bom(\square)}(x,y) )   \\
\leq  & \cO(1)  e_k C L^{(k-i)/2}  p(e_i) \   d'(x,y)^{-2} \exp( - \cO(1) d_{\bom(\square)}(x,y) )  \\
= & \cO(1)C L^{k-i} e_i p(e_i)\    d'(x,y)^{-2} \exp( - \cO(1) d_{\bom(\square)}(x,y) ) \\
\end{split}
\end{equation}
The second term has the form for   $ x \in \de \Om^{(k)}_j(\square)   \subset   L^{-k}\Om_0(\square)$, 
  $j \leq i+1$:
\begin{equation} 
\begin{split}
& (v^Q_{\square} (\cA,0)   S^*_{\square} (0))(x,y) \\
=  & 
 L^{2(k-j)} b_j\int_{|x'-[x]| \leq L^{-(k-j)}/2 }\left( \exp(ie_j \cA_{L^{k-j}}(\tilde \Ga_{z,[z]} \cup
\tilde \Ga_{[z],z'})-1\right)|_{\stackrel{z=L^{k-j}x}{z'=L^{k-j}x'}}
 (S^*_{\square}(0))(x',y) 
\end{split}
\end{equation}
The contour has length bounded by one  and so we have  
bound
\begin{equation}
|\left( \exp(ie_j \cA_{L^{k-j}}(\dots)-1\right) |  \leq    e_j  \sup | \cA_{L^{k-j}}|
\leq   C L^{-(i-j)/2}e_j p(e_i)  \leq   \cO(1)Ce_i p(e_i)
\end{equation}
and hence 
\begin{equation} 
\begin{split}
|(v^Q_{\square} (\cA,0)   S^*_{\square} (0))(x,y)| 
\leq    & 
 \cO(1)C L^{2(k-j)}e_i p(e_i)   \int_{|x'-[x]| \leq L^{-(k-j)}/2 } 
 d'(x,y)^{-2} \exp( - \cO(1)  d_{\bom(\square)}(x,y)) \\
\leq    & 
 \cO(1)C L^{k-j}e_i p(e_i)   \exp( - \cO(1)  d_{\bom(\square)}(x,y)) \\
\end{split}
\end{equation}
Combining the two bounds we have 
for  $x \in \de \Om^{(k)}_j(\square) $
\begin{equation}  \label{estimate2}
|(v_{ \square}(\cA,0)S^*_{\square}(0))(x,y)|
\leq   
 \cO(1)CL^{k-j} e_i p(e_i)     d'(x,y)^{-2}\exp( - \cO(1) d_{\bom(\square)}(x,y) ) 
\end{equation}

Now we can estimate the expansion.
Dividing  $\de \Om^{(k)}_j(\square)  $ into blocks  $\De$ of size  $L^{-(k-j)}$
and hence   $L^{-k} \Om_0(\square)$ into blocks of various sizes 
we have  
\begin{equation}  
\begin{split}
S^*_{ \square}(\cA,x,y)  =& \sum_{n=0}^{\infty}\sum_{\De_1, \dots,\De_n}
  \int_{\De_1}dx_1 \dots  \int_{\De_n}dx_n
  \\ 
 &(  S^*_{\square}(0,x,x_1)
 (v_{\square}(\cA)  S^*_{\square}(0))(x_1,x_2)
\cdots 
( v_{\square}(\cA) S^*_{\square}(0))(x_n,y)\\
\end{split}
\end{equation} 
We use our estimate   (\ref{estimate2})  as well as (\ref{est10}) and (\ref{est11})
and  find
\begin{equation}  
\begin{split}
|S^*_{ \square}(\cA,x,y)|   \leq 
 &   \left( \sum_{n=0}^{\infty} (\cO(1)Ce_ip(e_i) )^n \right) \ 
  d'(x,y)^{-2} \exp( - \cO(1)  d_{\bom(\square)}(x,y)) 
\\
\leq &  \cO(1)  d'(x,y)^{-2} \exp( - \cO(1)  d_{\bom(\square)}(x,y))
\\
\end{split}
\end{equation}
Here we use that  $e_i < e_k$ is assumed sufficiently small.   By 
(\ref{so}) we can replace  $\exp( - \cO(1)  d_{\bom(\square)}(x,y)) $ by  $\exp( - \cO(1)  d_{\bla}(x,y)) $
 to complete the proof in this case.
\bigskip

\noindent
\textbf{part V.}
Finally we extend the result to  the case  $|\pa \cA | 
\leq C L^{3(k-i)/2}p(e_i)$  on $\de\La_i^{(k)}$, and hence the same 
bound on   $ L^{-k}\Om_0(\square)$.
Instead of  (\ref{monday}) we use  
\begin{equation}  \label{tuesday}
S^*_{ \square}(\cA) =S^*_{ \square}(\bar \cA)
\left( \sum_{n=0}^{\infty} \left(  v_{ \square}(\cA,\bar \cA)S^*_{ \square}(\bar\cA)
\right)^n  \right)
\end{equation}
 where 
$\bar \cA$ is the average of $\cA$ over  $L^{-k} \Om_0(\square)$.
 Now  $\bar \cA$ is  pure gauge  and can be written  $\bar \cA  =  d   \la$.   and
Since  the propagator is  gauge invariant  we have
\begin{equation}
S^*_{ \square}( \bar \cA,x,y)   =  e^{-ie_k\la(x)} S^*_{ \square}(0,x,y)    e^{-ie_k\la(y)}
\end{equation}
which  satisfies the same bound  (\ref{same}) as   $S^*_{ \square}(0,x,y)$.
Also  
\begin{equation}
|\cA   -\bar \cA|  \leq 11r_1 M_0 L^{-(k-i)}  \sup | \pa  \cA |
 \leq C' L^{(k-i)/2}  C p(e_i)
\end{equation}
for a new constant  $C'$.
This is   a bound  of the form  we assumed on the field  in part IV.
One can then show that 
 $ v_{ \square}(\cA,   \bar  \cA)S^*_{ \square}(\bar \cA)$ satisfies the same bound 
 (\ref{estimate2})  used in part IV.
Thus we can repeat part IV with the same result.  This completes the proof of the lemma and the theorem.
\bigskip

As a corollary  we consider perturbation by a complex background field  $\cA'$.  We need
a bound on  $\cA'$ itself, not just  $\pa \cA'$.

\begin{cor}
Let   $\cA$ be real and   satisfy  $|\pa  \cA|  \leq  C L^{3(k-i)/2} p(e_i)$   on   $\de \La_i^{(k)}$  
and let   $\cA'$ be complex and    satisfy  $|\cA'|  \leq C  L^{(k-i)/2} p(e_i)$   on   $\de \La_i^{(k)}$.
Then   $\cS^*_{\square} (\cA + \cA')$    and   
 $S_{k, \bla, \om   }(\cA + \cA')$  and  $S_{k, \bla } (\cA + \cA' )$ 
all exist,  are analytic in  $\cA'$,  and satisfy the $\cA'=0$  bounds (\ref{a}),(\ref{b}), (\ref{star}),
now with
larger constants.
\end{cor}

\pr   It suffices to prove the result for   $\cS^*_{\square} (\cA + \cA')$, the others follow.
We make the expansion 
\begin{equation}  
S^*_{\square}(\cA + \cA') =S^*_{ \square}( \cA)
\left( \sum_{n=0}^{\infty} \left(  v_{ \square}(\cA+ \cA',\cA)S^*_{ \square}(\cA)
\right)^n   \right)
\end{equation}
The bound on $\pa \cA$ gives control over  $S_{\square}^*(\cA)$ by the theorem and the bound on complex $\cA'$
can be used to show that  
  $ v_{ \square}(\cA+ \cA',\cA)S^*_{ \square}(\cA)$ satisfies (\ref{estimate2}).
Here we use also  
\begin{equation}
|\exp(ie_kL^{-k}  \cA')|  \leq   \exp(e_kL^{-k}(CL^{(k-i)/2}p(e_i)) \leq  \exp(Ce_ip(e_i)) \leq 2
\end{equation}
Now repeat part IV of the lemma and  get the result. 
\bigskip

\subsection{bosons}

The treatment for bosons is similar, but easier since there is no background field.
The random walk expansion for the boson propagator has the form
\begin{equation}    \label{bwalk}
G_{k, \bla} (x,y )=   \sum_{\om: x \to y}  G_{k, \bla, \om }(x,y) 
\end{equation}
\bigskip

\begin{thm} Let   $M_0$  be   sufficiently large.
Then  
$G_{k, \bla}$ exists and has the  random walk expansion   (\ref{bwalk}).
We have 
\begin{equation}
\begin{split}
| G_{k,\bla,\omega}( x,y)  |
\leq & \cO(1) (\cO(1)M_0^{-1})^{|\om|} d'(x,y)^{-1} \exp( - \cO(1) d_{\bla}(x,y) ) \\
|G_{k, \bla } (x,y )|
\leq &  \cO(1) d'(x,y)^{-1}  \exp( - \cO(1) d_{\bla}(x,y) ) \\
\end{split}
\end{equation}
If  $x \in  \de \La_i^{(k)}$ then
\begin{equation}
\begin{split}
|\pa  G_{k, \bla , \om } (x,y ) |
\leq  & \cO(1) (\cO(1)M_0^{-1})^{|\om|}(L^{k-i} d'(x,y)^{-1}+  d'(x,y)^{-2})   \exp( - \cO(1) d_{\bla}(x,y) ) \\
|\pa  G_{k, \bla } (x,y ) |
\leq  &   \cO(1)(L^{k-i} d'(x,y)^{-1}+  d'(x,y)^{-2})   \exp( - \cO(1) d_{\bla}(x,y) ) \\
\end{split}
\end{equation}
\end{thm}
\bigskip

The proof depends on the lemma:
\bigskip

\begin{lem} 
\label{l}
Under the same hypotheses for each $\bbD_i$ block     $\square$ there is 
an operator     $G^*_{\square}$ such that   for   
$x \in \tilde  \square$:
\begin{equation}
( \left(  -\De+ \mu^2_k  +   \cQ_{k,\bla}^T  \bfa   \cQ_{k,\bla} \right) G^*_{\square}f)(x)  = f(x) 
\end{equation}
and  for  $x,y   \in \tilde  \square$
\begin{equation}
\begin{split}
|G^*_{\square}(x,y)| \leq   &  \cO(1)  d'(x,y)^{-1}\exp( - \cO(1) d_{\bla}(x,y) )  \\
 | \pa G^*_{\square}(x,y)| \leq &    \cO(1)  (L^{k-i} d'(x,y)^{-1}+  d'(x,y)^{-2}) ) \exp( - \cO(1) d_{\bla}(x,y) )  \\
\end{split}
\end{equation}
\end{lem}
\bigskip

Assuming the lemma we prove the theorem.
\bigskip

\pr The random walk expansion has the form   
\begin{equation}  \label{brandom}
\begin{split}
G_{k, \bla} =& \sum_{n=0}^{\infty}  \sum_{ \square_0, \square_1,...,\square_n}
( h_{\square_0} G^*_{\square_{0}}    h_{\square_{0}})
(R_{\square_1}    G^*_{\square_{1}}    h_{\square_{1}})
\cdots 
(R_{\square_n}    G^*_{\square_{n}}    h_{\square_{n}})\\
\equiv & \sum_{\om}   G_{k,\bla,\omega}  \\
\end{split}
\end{equation}
where now
\begin{equation}
 R_{\square} = - \left[ ( -\De+ \mu^2_k  +   \cQ_{k,\bla}^T  \bfa   \cQ_{k,\bla}), h_{\square}\right]
\end{equation}

We   write    $R_{\square} =  R^{\De}_{\square} +  R^Q_{\square}$ and    estimate  for  $\square \in \bbD_i$
\begin{equation}
\begin{split}
|(R^{\De}_{\square} G^*_{\square})(x,y)|  =&  | (-\De h_{\square})(x) G^*_{\square}(x,y)
+  (\pa h_{\square})(x)  \pa G^*_{\square}(x,y)+ (\pa^T h_{\square})(x)  \pa^T G^*_{\square}(x,y)|  \\
\leq  & \cO(1) M_0^{-1} \left(  L^{2(k-i)}   d'(x,y)^{-1}
+  L^{k-i}   d'(x,y)^{-2}  \right)
\exp( - \cO(1) d_{\bla}(x,y) ) \\
\end{split}
\end{equation}
The same bound holds easily for   $|(R^Q_{\square}G^*_{\square})(x,y)|$   and hence it holds also for  
$|(R_{\square}G^*_{\square})(x,y)|$.

Now we follow the proof of theorem  \ref{f}.  The only difference is in the  short distance estimates
which  we modify 
as follows .   Instead of  (\ref{est10})
we have    by (\ref{e1}), (\ref{e2}), (\ref{e3})
\begin{equation}
\begin{split}
&\int_{\De_j}  \left( L^{2(k-i_j)}   d'(x_{j-1},x_j)^{-1}
+  L^{k-i_j}   d'(x_{j-1},x_{j})^{-2}  \right) 
\\
&
\ \ \ \ \ \left( L^{2(k-i_j)}   d'(x_j,x_{j+1})^{-1}
+  L^{k-i_j}   d'(x_j,x_{j+1})^{-2}  \right)
dx_j \\
& \  \leq \cO  (L^{2(k-i_j)})  d'(x_{j-1},x_{j+1})^{-1}
+  \cO(L^{k-i_j})   d'(x_{j-1},x_{j+1})^{-2}   
\end{split}
\end{equation}
As we repeat this estimate we have   to adjust the  $i$ in the factor  $L^{k-i}$ so that neighbors match.
But since  $|i_j-i_{j+1}| \leq 1$ this  costs
at  most  $\cO(L^{2n})$ which we can afford.
In the last step  since  $|i-i_0| \leq 1$  the inequality  is  
\begin{equation}
 \int_{\De_1}   d'(x,x_1)^{-1}
\left(  \cO(L^{2(k-i)})   d'(x_1,y)^{-1}
+  \cO(L^{k-i} ) d'(x_1,y)^{-2}  \right) dx_1  
\leq   \cO(1)  d'(x,y)^{-1}
\end{equation}
The rest of the proof is as before and gives the result for  $G_{k, \bla, \om}$ and  $ G_{k,\bla}$.
For the bounds on    $\pa G_{k, \bla, \om}$ and  $\pa G_{k,\bla}$ the last step is  
\begin{equation}
\begin{split}
&\int_{\De_1}  \left( L^{k-i}   d'(x,x_1)^{-1}
+   d'(x,x_1)^{-2}  \right) 
\left( L^{2(k-i)}   d'(x_1,y)^{-1}
+  L^{k-i}   d'(x_1,y)^{-2}  \right)
dx_1 \\
& \  \leq \cO(1) (  L^{k-i}  d'(x,y)^{-1}
+     d'(x,y)^{-2})   
\end{split}
\end{equation}
to complete the proof.
\bigskip

The lemma is proved in much the same way.  One follows  parts  I-III of lemma \ref{l} modifying the short distance 
behavior as indicated above.

\newpage

\appendix

\section{Averaging operators}  \label{A}

\subsection{bosons}

We consider the $k$-step boson averaging operator defined recursively in  (\ref{bQ})  by
 $\cQ_0 = \textrm{id}$ and
\begin{equation} 
\cQ_{k+1}
=  \si_{L^{-1}}   Q \ \cQ_{k}  \ \si_{L}
\end{equation}
Then we have  
\begin{lem}
\begin{equation} \label{q1}
\begin{split}
&\int  d A_k \  \cN_{k+1,a}^{-1}     \cN_{k,a_k}^{-1}     
 \exp \left( - \frac{a}{2L^2}   |A_{k+1,L}- Q A_k|^2  \right) 
 \exp \left( -  \frac{a_k}{2}   |A_{k}- \cQ_{k} \cA_L|^2  \right)  \\
\ \ \ \ \ \ \ \ = &   \cN_{k+1,a_{k+1}}^{-1}    \exp \left(
 - \frac{a_{k+1}}{2}   |A_{k+1}- \cQ_{k+1} \cA|^2  
\right) \\
\end{split}
\end{equation}
\end{lem}
\bigskip

\pr
 Shift  the integration variable   $A_k \to A_k +  \cQ_{k} \cA_L$, identify  $\cQ_{k+1}$, and then
the left side can be written with   $\Phi  = A_{k+1}- \cQ_{k+1} \cA$
  \begin{equation}
\begin{split}
&\int  d A_k \  \cN_{k+1,a}^{-1}     \cN_{k,a_k}^{-1}     
 \exp \left( - \frac{a}{2L^2}   |\Phi_L- Q A_k|^2  \right) 
 \exp \left( -  \frac{a_k}{2}   |A_{k}|^2  \right)\\
=&
\const \exp \left( - \frac{a}{2}   |\Phi|^2  \right) 
\int  d A_k   
 \exp \left( - \frac{a}{L^2}   (Q^T\Phi_L, A_k)        \right) 
 \exp \left( - \frac{1}{2}   (A_k, (    a_k + \frac{a}{L^2}Q^TQ) A_k)   \right) \\
=&
\const \exp \left( - \frac{a}{2}   |\Phi|^2  \right) 
\exp  \left(\frac{a^2}{2L^4} (Q^T\Phi_L, (    a_k + \frac{a}{L^2}Q^TQ) ^{-1} Q^T\Phi_L)\right)\\
=&
\const \exp \left( - \frac{a}{2}   |\Phi|^2  \right) 
\exp  \left(  \frac{a^2}{2L^2} (\Phi, (    a_k + \frac{a}{L^2}) ^{-1} \Phi)\right)\\
=&
\const \exp \left( - \frac{a_{k+1}}{2}   |\Phi|^2  \right) 
\\
\end{split}
\end{equation}
Here we have used  $QQ^T = I$ and  $a_{k+1}= aa_k/(a_k + aL^{-2})$ .  Since the left side integrates to one,
the right side  integrates to one which means the constant must be   $\cN_{k+1,a_{k+1}}^{-1}$ and
we have the result.
\bigskip

We also use a local variation of this result.  Let  $\La  \subset  \bbT^0_{N+M-k-1}$  so  $L\La \subset 
 \bbT^1_{N+M-k}$  and the blocked set  $B_1 \La \subset  \bbT^0_{N+M-k}$.  Then with  $\cN_{\La,a}
= ( 2\pi/a)^{3| \La|/2}$ we have 
\begin{equation}  \label{q2}
\begin{split}
&\int  d A_{k,B_1\La} \  \cN_{\La,a}^{-1}     \cN_{B_1\La ,a_k}^{-1}     
 \exp \left( - \frac{a}{2L^2}   |A_{k+1,L}- Q A_k|^2_{L\La}  \right) 
 \exp \left( -  \frac{a_k}{2}   |A_{k}- \cQ_{k} \cA_L|^2_{B_1\La}  \right)  \\
\ \ \ \ \ \ \ \ = &   \cN_{\La,a_{k+1}}^{-1}    \exp \left(
 - \frac{a_{k+1}}{2}   |A_{k+1}- \cQ_{k+1} \cA|^2_{\La}  
\right) \\
\end{split}
\end{equation}

\subsection{fermions}

For fermions the multiple  averaging
$ \cQ_{k}(a_{k-1}, \dots , a_{0})$ 
depend on fields   $a_{k-1}, \dots , a_{0}$ all on   $\bbT^{-k}_{N+M-k}$.
and  are defined recursively in  (\ref{together}) by   $\cQ_0 = \textrm{id}$ and  
\begin{equation}   
\cQ_{k+1}(a_{k}, \dots , a_0) 
=  \si_{L^{-1}}\   Q_{e_k}(\tilde \cQ a_{k,L})\
 \cQ_{k}(a_{k-1,L}, \dots , a_{0,L})  \ \si_{L}
\end{equation}
Then we have   

\begin{lem}
\begin{equation}  
\begin{split}
 &\int  d \Psi_k  \cM_{k+1,b}^{-1}     \cM_{k,b_k}^{-1} 
 \exp \left(
 - \frac{b}{L}   |\Psi_{k+1,L}- Q_{e_k}( \tilde \cQ_k a_{k,L}) \Psi_k|^2  
\right) \\
&\ \ \ \ \ \ \ \ \ \ \
 \exp \left(
 - b_k   |\Psi_{k}- \cQ_{k}(a_{k-1,L}, \dots , a_{0,L}) \psi_L|^2  
\right)  \\
&=\cM_{k+1,b_{k+1}}^{-1}  \exp \left(
 - b_{k+1}   |\Psi_{k+1}- \cQ_{k+1}(a_k, \dots , a_0) \psi|^2  
\right) \\ 
\end{split}
\end{equation}
\end{lem}
\bigskip

\re   If we take   $a_{j,L} =  (\cA_j)_{L^{-(k-j)}} $ for  $\cA_j$ on   $\bbT^{-j}_{m+N-j}$
then we recover the version (\ref{newt}) quoted in the text .
\bigskip

\pr
First   shift   $\Psi_k  \to \Psi_{k}+ \cQ_{k}(a_{k-1,L}, \dots , a_{0,L})\psi_L$
and    $\bPsi_k  \to \bPsi_{k}+ \cQ_{k}(-a_{k-1,L}, \dots , -a_{0,L})\bpsi_L$
and identify  $ \cQ_{k+1}(a_{k}, \dots , a_0)$.
Then  left side can be written with   $\Phi  = \Psi_{k+1}- \cQ_{k+1}(a_{k}, \dots , a_0)\psi$
and   $\bar \Phi  = \bPsi_{k+1}- \cQ_{k+1}(-a_{k}, \dots , -a_0)\bpsi$   and  $A =   \tilde \cQ_k a_{k,L}$    : 
\begin{equation}  
\begin{split}
 &\int  d \Psi_k  \cM_{k+1,b}^{-1}     \cM_{k,b_k}^{-1} 
 \exp \left(
 - \frac{b}{L}   |\Phi_{L}- Q_{e_k}(A) \Psi_k|^2  
\right) 
 \exp \left(
 - b_k   (\bPsi_{k}, \Psi_k) 
\right)  \\
= &\const   \exp(-b (\bar \Phi,\Phi))  \int  d \Psi_k   
 \exp(  - \frac{b}{L}   (Q_{e_k}(A)^T \bar \Phi_L,\Psi_k))
 \exp(  - \frac{b}{L}  (\bPsi_k, Q_{e_k}( -A)^T  \Phi_L))
\\ 
& \exp(  -  (\bPsi_k,  [ b_k +  \frac{b}{L}   Q_{e_k}( -A)^T  Q_{e_k}(A)]  \Psi_k))
\\
=& \const   \exp(-b (\bar \Phi,\Phi))   
 \exp(   \frac{b^2}{L^2}  (Q_{e_k}(A)^T \bar \Phi_L,
[ b_k +    \frac{b}{L}   Q_{e_k}( -A)^T  Q_{e_k}(A)]^{-1} Q_{e_k}( -A)^T  \Phi_L))
\\ 
=& \const   \exp(-b (\bar \Phi,\Phi))    
 \exp(   \frac{b^2}{L}  (\bar \Phi,( b_k +   \frac{b}{L})^{-1}   \Phi))
\\ 
=& \const   \exp(- b_{k+1}(\bar \Phi,\Phi))    
\\ 
\end{split}
\end{equation}
Here we have used  $Q_{e_k}(A) Q_{e_k}(-A) ^T = I$ and   $b_{k+1}= bb_k/(b_k + bL^{-1})$.  Since the
left side integrates to one, the right side  integrates to one which means the constant must be  
$\cM_{k+1,b_{k+1}}^{-1}$ and we have the result.
\bigskip

We also want to work out an explicit expression for  $\cQ_k(a_{k-1}, \dots, a_0)$.
First for   $k=1$ we consider    
$
\cQ_1(a_0)  =  \si_L^{-1}  Q_{e_0}(a_{0,L}) \si_L
$
and  compute  it   as 
\begin{equation}
\begin{split}
((\cQ_1(a_0) \psi)(y)  =&    
L^{-3}  \sum_{ | x- Ly| < L/2}  \exp \left( ie_0 a_{0,L}(\Ga_{Ly,x})\right) \psi(x/L)\\
 =&    
  \int_{ | x- y| < 1/2}  \exp \left( ie_0 a_{0,L}(\Ga_{Ly,Lx})\right) \psi(x)\\
 =&    
  \int_{ | x- y| < 1/2}  \exp \left( ie_1 a_0( \Ga_{y,x})\right)
\psi(x)\\
\end{split} 
\end{equation}  \label{gscale}
Here we have used  $e_1 = L^{1/2}e_0$ and  
\begin{equation}   \label{forty}
\cA_L(\Ga_L)  = L^{1/2}  \cA(\Ga) 
\end{equation}
\bigskip

For the general case we need some definitions. Recall that 
  $ \tilde \cQ_j$ maps from functions on  $\bbT^{-j}_{N+M-j}$ to   $\bbT^{0}_{N+M-j}$
For   $j \leq k  $ we also consider  the scaled version
 $\si_{L^{-(k-j)} }\tilde \cQ_j   \si_{L^{k-j} }$
which is a map from functions on  $\bbT^{-k}_{N+M-k}$ to   $\bbT^{-(k-j)}_{N+M-k}$
and is also denoted $ \tilde \cQ_j$.
Also 
given  $y \in  \bbT^{0}_{N+M-k}$ and      $x \in \bbT^{-k}_{N+M-k}$
with  $|x-y| \leq 1/2$  we define a sequence of
points
$x_0 = x, x_1, \dots, x_{k-1}, x_k=y$  where $x_j$ is the    point in    $\bbT^{-(k-j)}_{N+M-k}$
such that   $|x-x_j| <  L^{-{(k-j)}}/2$.  Finally let  $ \Ga_{x_{j+1},x_j}$ be the standard contour 
in   $\bbT^{-(k-j)}_{N+M-k}$ taking $x_j $ to $x_{j+1}$.

\begin{lem}
\begin{equation}
\begin{split}
& (\cQ_k(a_{k-1}, \cdots, a_{0} )\psi )(y)
=   \int_{ | x- y| < 1/2}  \exp \left(
 ie_k\sum_{j=0}^{k-1}  (\tilde \cQ_j  a_j)( \Ga_{x_{j+1},x_j}) 
\right) \psi(x)\\
\end{split}
\end{equation}
\end{lem}
\bigskip

\pr   By induction. We know it is true for  $k=1$ and  assuming it is true for $k$ we compute
\begin{equation}
\begin{split}
&
(\cQ_{k+1}(a_{k}, \cdots,  a_{0})\psi)(y')= \\
=&L^{-3} \int_{ | x- L y'| < L/2}\exp(ie_k (\tilde \cQ_ka_{k,L})(\Ga_{Ly',x_k}) ) 
 \exp \left(
 ie_k\sum_{j=0}^{k-1}   (\tilde \cQ_j  a_{j,L})( \Ga_{x_{j+1},x_j})  
\right)
\psi(x/L)\\
\end{split}
\end{equation}
Now  make the change of variables  $x = Lx'$. Then   $x_j = Lx'_j$ 
and we  obtain
\begin{equation}
\begin{split}
&
(\cQ_{k+1}(a_{k}, \cdots,  a_{0})\psi)(y')= \\
=& \int_{ | x'-  y'| < 1/2}\exp  \left(ie_k(\tilde \cQ_ka_{k,L})(\Ga_{Ly',Lx'_k})  
 +
 ie_k\sum_{j=0}^{k-1}  (\tilde \cQ_j  a_{j,L})( \Ga_{Lx'_{j+1},Lx'_{j}})  
\right)
\psi(x')\\
=& \int_{ | x'-  y'| < 1/2}
 \exp \left(
 ie_k\sum_{j=0}^{k} (\tilde \cQ_j  a_{j,L})( \Ga_{Lx'_{j+1},Lx'_{j}})  
\right)
\psi(x')\\
\end{split}
\end{equation}
with $x'_{j+1} = y'$.
Now we use   (\ref{forty})   and  $ e_{k+1} =L^{1/2}e_k$    
to write this as   
\begin{equation}
 (\cQ_{k+1}(a_{k}, \cdots, a_{0} )\psi )(y')
=   \int_{ | x'- y'| < 1/2}  \exp \left(
 ie_{k+1}\sum_{j=0}^{k}  (\tilde \cQ_j  a_j)( \Ga_{x'_{j+1},x'_j}) 
\right) \psi(x')
\end{equation}
which is the statement for  $k+1$.

\section{Perturbation identities}  \label{B}

Our goal is to prove the identity  (\ref{basic}).  We start with
the recursion relation   (\ref{winter}) for  $\rho^{\star}_{k}(t,\Psi_{k}, \cA_{k})$ and introduce under
the  integral sign  the characteristic function
\begin{equation}
\chi\left( \frac{A_k}{p(te_k)}\right)= \prod_{x, \mu }\chi\left( \frac{A_{k,\mu}(x)}{p(te_k)}\right)
\end{equation}
Here $\chi$ is a smooth functions  satisfying  $\chi = 1$ on  $[-1/2,1/2]$ and  $\chi = 0$ outside 
$[-1,1]$, and  $p(e) = [ \log(e^{-1})]^p$.  This is a continuous function for    $t\geq 0$ if we set
  $\chi( A_{\mu}(x)/p(te_k))|_{t=0} =1$.  
The new recursion relation defines  new functions which we call  $\sigma^{\star}_{k}(t,\Psi_{k}, \cA_{k})$.
Thus we have 
\begin{equation}  \label{winter2}
\begin{split}
& \sigma^{\star}_{k+1}(t,\Psi_{k+1}, \cA_{k+1})=
\int  d \Psi_k \ d\mu_{C_k}(A_k)\      \chi\left( \frac{A_k}{p(te_k)}\right)  \cM_{k+1,b}^{-1}  \\
&
\exp \left(-   \frac{b}{L}   | \Psi_{k+1,L} -
Q_{e_k}(\tilde \cQ_k ( \cA_{k+1,L} +
t\cA_k))\Psi_k|^2
\right) \sigma^{\star}_{k}(t,\Psi_k,  \cA_{k+1,L} +t \cA_k)\\
\end{split}
\end{equation}
The starting function for  $k=0$ is again
$ \exp (  -   (\Psi_0 ,( D_{e_0} ( \cA_0 )
+ m_0) \Psi_0 ) )$.   One can show inductively that  $\sigma^{\star}_{k}(t,\Psi_{k}, \cA_{k})$
is a bounded analytic function of  $\cA_k$ on a neighborhood of the real axis, and that
it is a continuous function of $t\geq 0$, smooth for  $t>0$.

\begin{lem}  The derivatives of  $\sigma^{\star}_{k}(t,\Psi_{k}, \cA_{k})$ at $t=0$ exist 
and are equal to those of   $\rho^{\star}_{k}(t,\Psi_{k}, \cA_{k})$. 
\end{lem}
\bigskip

\pr (after \cite{BIJ88})  We suppose it is true for $k$ and establish it for  $k+1$.    
The point is that the derivatives    of  $ \chi( \cA_k/p(te_k))$
do not contribute, and we focus on this   aspect.

Consider the first derivative of  $\sigma^{\star}_{k+1}(t)$. Since it is
continuous at zero we can find the derivative at  zero by taking the limit from positive values.  The contribution 
from the characteristic function has the form    
\begin{equation}
\begin{split}
&\lim_{t \to 0^+}\int d \Psi_k \ d\mu_{C_k}(A_k)\
 \frac{d}{dt} \chi\left( \frac{A_k}{p(te_k)}\right)   \cM_{k+1,b}^{-1}  \\
&
\exp \left(-   \frac{b}{L}   | \Psi_{k+1,L} -
Q_{e_k}(\tilde \cQ_k ( \cA_{k+1,L} +
t\cA_k))\Psi_k|^2
\right) \sigma^{\star}_{k}(t,\Psi_k,  \cA_{k+1,L} +t \cA_k)\\
& =  \lim_{t \to 0^+}\int  \ d\mu_{C_k}(A_k)\
 \frac{d}{dt} \chi( \frac{A_k}{p(te_k)}) f(t,\cA_k)
\end{split}
\end{equation}
where  $f(t,\cA_k)$ is a bounded continuous function.  We must show that this limit is zero.

Now 
\begin{equation}
\frac{d}{dt} \chi\left( \frac{A_{k, \mu}(x)}{p(te_k)}\right) = \chi'\left( \frac{A_{k,\mu}(x)}{p(te_k)}\right)
\frac{A_{k,\mu}(x)}{ p(te_k)} \frac{-pt^{-1}}{ p(te_k)^{ 1/p}}
\end{equation}
is $\cO(t^{-1})$ which is not sufficient by itself.  However the 
integrand vanishes if     $|A_{k,\mu}(x)|< p(te_k)/2$  for all  $x,\mu$.
Thus the expression is dominated by  
\begin{equation}
\cO(t^{-1})\sum_{x,\mu}   \int_ {|A_{k,\mu}(x)|\geq  p(te_k)/2}    d\mu_{C_k}(A_k)\ 
\leq \cO(t^{-1} e^{- p(te_k)})  \to 0 
\end{equation}
Higher derivatives can be treated in the same way.
\bigskip

Next   assuming   $e_k | \pa \cA_k|$ is sufficiently small we   may define
$\sigma^{\bullet}_{k}(t,\Psi_k,  \psi_k(\cA_k),  \cA_k) $ by   
\begin{equation}   \label{newbullet}
 \sigma^{\star}_k(t,\Psi_k, \cA_k) 
=  Z_k(\cA_k)\exp(- (\bPsi_k, D_k(\cA_k) \Psi_k) )\sigma^{\bullet}_{k}(t,\Psi_k,  \psi_k(\cA_k),  \cA_k) 
\end{equation}
Then   $\sigma^{\bullet}_{k}(t,\Psi_k,  \psi_k(\cA_k),  \cA_k)$ and $\rho^{\bullet}_{k}(t,\Psi_k, 
\psi_k(\cA_k),  \cA_k)$ have the same derivatives at $t=0$.
In particular  
 $ (\sigma_k^{\bullet})''(0)/2 =  (\rho_k^{\bullet})''(0)/2 =  P_k$.

Now we can prove  (\ref{basic}) which we repeat:

\begin{lem}  Let  $e_{k+1} | \pa \cA_{k+1}|$  be sufficiently small.  Then
 \begin{equation} 
\begin{split}
&  P^+_{k+1}(\Psi_{k+1},\psi_{k+1}(\cA_{k+1}),\cA_{k+1}) \\
= &  \left[  P^+_k( \Psi(\cA)  ,[ \psi_{k+1}(\cA_{k+1})]_L,\ \cA)
-  (\De ^{\Ga}_{ \Psi_k}P_{k} )^+  ( \Psi(\cA)  ,[ \psi_{k+1}(\cA_{k+1})]_L,\ \cA) \right.
 \\  
+  & \frac12   \int_{z,w}  J_{k}(z)
\tilde   C_k(z,w)  J_{k}(w)
 -\frac12 \int_{z,w}   J_{k} (z,w)
\tilde  C_k(z,w) \\
 - & \left. \int_{z,w}   \sum_{x,y}    \bar  K_{k}(z,x)
\Ga_k ( \cA;x,y)   K_{k}(w,y) \tilde  C_k(z,w)
 \right]_{\stackrel{\Psi =\Psi_{k+1,L}}{ \cA =  \cA_{k+1,L}}} \\
\end{split}
\end{equation}
\end{lem}
\bigskip

\pr  In   (\ref{winter2}) insert the expression
  (\ref{newbullet}) for  $  \sigma^{\star}_{k}(t,\Psi_k,   \cA_{k+1,L} +t\cA_k) $.
This  representation is  possible because   $e_k|\pa( \cA_{k+1,L} +t \cA_k)|$ 
is small. The first term is small  by our  assumption and the second term    
is small by the characteristic function since 
\begin{equation}
te_k|  \pa \cA_k| =te_k|  \pa \cH_k A_k| \leq  \cO(te_k) \sup| A_k|  \leq  \cO(    te_k p(te_k))
\end{equation}  
Next we   carry out the steps in section \ref{ss}.  These were formal  for  $\rho_k^{\bullet}$ but
are now rigorous.  This yields instead of (\ref{spring})
\begin{equation}
\begin{split}
 & 
\sigma^{\bullet}_{k+1}(t,\Psi_{k+1},\psi_{k+1}(\cA_{k+1}), \cA_{k+1}) \\
=&
\int  d \mu_{\Ga_k(\cA)}(\Psi_k)\ d \mu_{C_k}(A_k)     \chi \left( \frac{\cA_k}{p(te_k)}\right) 
\exp \left( -V_k(\Psi,\Psi(\cA) + \Psi_k, \cA, t\cA_k)
 -U_k(\cA ,t\cA_k) \right)    \\
 &
 \sigma^{\bullet}_k(t,\Psi(\cA)  + \Psi_k, [ \psi_{k+1}(\cA_{k+1})]_L  +\psi_k(\cA)+
\de \cH_k(\cA,t \cA_k)(\Psi(\cA)  + \Psi_k),\ 
\cA+t\cA_k    )
 |_{\stackrel{\Psi=\Psi_{k+1,L}}{ \cA = \cA_{k+1,L}}} \\
\end{split}
\end{equation}
Now take two derivatives    at  $t=0$. The derivatives of  $ \chi( \cA_k/p(te_k))$
do not contribute as we have explained. The derivatives   of   $\sigma_k^{\bullet},
\sigma_{k+1}^{\bullet}$  give  $P_k, P_{k+1}$,  and the derivatives of  $V_k$ give $J_k,K_k$.
 The details are  explained in the text.

\end{document}